\documentclass[sigconf, screen]{acmart}

\AtBeginDocument{%
  \providecommand\BibTeX{{%
    Bib\TeX}}}

\setcopyright{none}
\renewcommand\footnotetextcopyrightpermission[1]{}
\acmConference[MICRO 2025]{The 58th IEEE/ACM International Symposium on Microarchitecture}{October 18--22, 2025}{Seoul, Korea}

\usepackage{multirow}
\usepackage{comment}
\usepackage{verbatim} 

\usepackage{tikz}
\usepackage{pgfplots}
\usepackage{pgfplotstable}
\usepackage{graphicx}
\usepackage{multirow}
\usepackage{pifont}
\usepackage[export]{adjustbox}
\usepackage{pgf-pie}
\usepackage{tkz-graph}
\usepackage{graphicx}
\usepackage{subcaption}
\usepackage[skip=2pt]{caption}
\usepackage{hyperref}
\usepackage{xcolor}
\hypersetup{
    colorlinks,
    linkcolor={red!40!black},
    citecolor={blue!40!black},
    urlcolor={black}
}
\usepackage{algorithm}
\usepackage{algpseudocode}
\usepackage{amsmath}

\usepackage{booktabs}
\usepackage{titlesec}
\usepackage{placeins}
\usepackage{float}
\usepackage{dblfloatfix}
\usepackage{makecell}

\usetikzlibrary{pgfplots.groupplots}
\usetikzlibrary{shadows,patterns,shapes,arrows,decorations.pathmorphing,backgrounds,positioning,fit,plotmarks,calc,spy,matrix}
\pgfplotsset{compat=1.11,
    /pgfplots/ybar legend/.style={
    /pgfplots/legend image code/.code={%
       \draw[##1,/tikz/.cd,yshift=-0.25em]
        (0cm,0cm) rectangle (3pt,0.8em);},
   },
}
\usepackage{titlesec}
\usetikzlibrary{decorations.text}
\definecolor{cadetblue}{rgb}{0.37, 0.62, 0.63}
\definecolor{airforceblue}{rgb}{0.36, 0.54, 0.66}
\definecolor{caribbeangreen}{rgb}{0.0, 0.8, 0.6}
\definecolor{carolinablue}{rgb}{0.6, 0.73, 0.89}
\definecolor{darkgoldenrod}{rgb}{0.72, 0.53, 0.04}
\definecolor{debianred}{rgb}{0.84, 0.04, 0.33}
\definecolor{fuzzywuzzy}{rgb}{0.8, 0.4, 0.4}
\definecolor{grullo}{rgb}{0.66, 0.6, 0.53}
\definecolor{ceil}{rgb}{0.57, 0.63, 0.81}
\definecolor{candypink}{rgb}{0.89, 0.44, 0.48}
\definecolor{calpolypomonagreen}{rgb}{0.12, 0.3, 0.17}
\definecolor{burntsienna}{rgb}{0.91, 0.45, 0.32}
\definecolor{atomictangerine}{rgb}{1.0, 0.6, 0.4}
\definecolor{goldenrod}{rgb}{0.85, 0.65, 0.13}
\definecolor{gamboge}{rgb}{0.89, 0.61, 0.06}
\definecolor{amber}{rgb}{1.0, 0.75, 0.0}
\definecolor{battleshipgrey}{rgb}{0.52, 0.52, 0.51}
\definecolor{darkcerulean}{rgb}{0.03, 0.27, 0.49}
\definecolor{fuzzywuzzy}{rgb}{0.8, 0.4, 0.4}
\definecolor{mediumseagreen}{rgb}{0.24, 0.7, 0.44}
\definecolor{antiquebrass}{rgb}{0.8, 0.58, 0.46}
\definecolor{apricot}{rgb}{0.98, 0.81, 0.69}
\definecolor{asparagus}{rgb}{0.53, 0.66, 0.42}
\definecolor{bananamania}{rgb}{0.98, 0.91, 0.71}
\definecolor{cadmiumgreen}{rgb}{0.0, 0.42, 0.24}
\definecolor{chocolate}{rgb}{0.48, 0.25, 0.0}
\definecolor{cinereous}{rgb}{0.6, 0.51, 0.48}
\definecolor{aliceblue}{rgb}{0.94, 0.97, 1.0}
\definecolor{beaublue}{rgb}{0.74, 0.83, 0.9}
\definecolor{blizzardblue}{rgb}{0.67, 0.9, 0.93}
\definecolor{bittersweet}{rgb}{1.0, 0.44, 0.37}
\definecolor{camouflagegreen}{rgb}{0.47, 0.53, 0.42}
\definecolor{darkolivegreen}{rgb}{0.33, 0.42, 0.18}
\definecolor{darkpastelblue}{rgb}{0.47, 0.62, 0.8}
\definecolor{desertsand}{rgb}{0.93, 0.79, 0.69}
\definecolor{deeppeach}{rgb}{1.0, 0.8, 0.64}
\definecolor{indianred}{rgb}{0.8, 0.36, 0.36}
\definecolor{oldmauve}{rgb}{0.4, 0.19, 0.28}
\definecolor{lightblue}{rgb}{0.68, 0.85, 0.9}
\definecolor{lightcyan}{rgb}{0.88, 1.0, 1.0}
\definecolor{viridian}{rgb}{0.25, 0.51, 0.43}
\definecolor{slategray}{rgb}{0.44, 0.5, 0.56}
\definecolor{manatee}{rgb}{0.59, 0.6, 0.67}
\definecolor{darkbrown}{rgb}{0.4, 0.26, 0.13}
\definecolor{almond}{rgb}{0.94, 0.87, 0.8}
\def\BibTeX{{\rm B\kern-.05em{\sc i\kern-.025em b}\kern-.08em
    T\kern-.1667em\lower.7ex\hbox{E}\kern-.125emX}}
\usepackage{colortbl}
\usepackage{tabulary}
\pgfkeys{%
/piechartthreed/.cd,
scale/.code                =  {\def\piechartthreedscale{#1}},
mix color/.code            =  {\def\piechartthreedmixcolor{#1}},
background color/.code     =  {\def\piechartthreedbackcolor{#1}},
name/.code                 =  {\def\piechartthreedname{#1}}}
\newcommand\piechartthreed[2][]{%
   \pgfkeys{/piechartthreed/.cd,
     scale            = 1,
     mix color        = gray,
     background color = white,
     name             = pc} 
  \pgfqkeys{/piechartthreed}{#1}
  \begin{scope}[scale=\piechartthreedscale] 
  \begin{scope}[xscale=5,yscale=3] 
     \path[preaction={fill=black,opacity=.8,
         path fading=circle with fuzzy edge 20 percent,
         transform canvas={yshift=-15mm*\piechartthreedscale}}] (0,0) circle (1cm);
     \pgfmathsetmacro\totan{0} 
     \global\let\totan\totan 
     \pgfmathsetmacro\bottoman{180} \global\let\bottoman\bottoman 
     \pgfmathsetmacro\toptoman{0}   \global\let\toptoman\toptoman 
     \begin{scope}[draw=black,thin]
     \foreach \an/\col [count=\xi] in {#2}{%
     \def\space{ } 
        \coordinate (\piechartthreedname\space\xi) at (\totan+\an/2:0.75cm); 
        \ifdim 180pt>\totan pt 
         \ifdim 0pt=\toptoman pt
            \pgfmathsetmacro\toptoman{180} 
            \global\let\toptoman\toptoman         
            \else
          \fi
        \fi   
        \fill[\col!80!gray,draw=black] (0,0)--(\totan:1cm)  arc(\totan:\totan+\an:1cm)
                                     --cycle;     
       \pgfmathsetmacro\finan{\totan+\an}
       \ifdim 180pt<\finan pt 
         \ifdim 180pt=\bottoman pt
            \shadedraw[left color=\col!20!\piechartthreedmixcolor,
                       right color=\col!5!\piechartthreedmixcolor,
                       draw=black,very thin] (180:1cm) -- ++(0,-3mm) arc (180:\totan+\an:1cm) 
                                                       -- ++(0,3mm)  arc (\totan+\an:180:1cm);
            \pgfmathsetmacro\bottoman{0}
            \global\let\bottoman\bottoman
            \else
            \shadedraw[left color=\col!20!\piechartthreedmixcolor,
                       right color=\col!5!\piechartthreedmixcolor,
                       draw=black,very thin](\totan:1cm)-- ++(0,-3mm) arc(\totan:\totan+\an:1cm)
                                                        -- ++(0,3mm)  arc(\totan+\an:\totan:1cm); 
          \fi
        \fi
        \pgfmathsetmacro\totan{\totan+\an}  \global\let\totan\totan 
       } 
    \end{scope}
   \end{scope}  
\end{scope}
}

\usepackage{todonotes}

\author{Rohan Juneja}
\affiliation{%
  \institution{National University of Singapore}
  \country{}
}
\email{rohan@comp.nus.edu.sg}

\author{Pranav Dangi}
\affiliation{%
  \institution{National University of Singapore}
  \country{}
}
\email{dangi@comp.nus.edu.sg}

\author{Thilini Kaushalya Bandara}
\affiliation{%
  \institution{National University of Singapore}
  \country{}
}
\email{thilini@comp.nus.edu.sg}

\author{Tulika Mitra}
\affiliation{%
  \institution{National University of Singapore}
  \country{}
}
\email{tulika@comp.nus.edu.sg}

\author{Li-Shiuan Peh}
\affiliation{%
  \institution{National University of Singapore}
  \country{}
}
\email{peh@comp.nus.edu.sg}

\begin{document}

\title{Nexus Machine: An Active Message Inspired Reconfigurable Architecture for Irregular Workloads}




\begin{abstract}
    Modern reconfigurable architectures are increasingly favored for resource-constrained edge devices as they balance high performance, energy efficiency, and programmability well. 
    However, their proficiency in handling regular compute patterns constrains their effectiveness in executing irregular workloads, such as sparse linear algebra and graph analytics with unpredictable access patterns and control flow. To address this limitation, we introduce the \textit{Nexus Machine}, a novel reconfigurable architecture consisting of a PE array designed to efficiently handle irregularity by distributing sparse tensors across the fabric and employing active messages that morph instructions based on dynamic control flow. 
    As the inherent irregularity in workloads can lead to high load imbalance among different Processing Elements (PEs), \textit{Nexus Machine} deploys and executes instructions en-route on idle PEs at run-time. 
    Thus, unlike traditional reconfigurable architectures with only static instructions within each PE, \textit{Nexus Machine} brings dynamic control to the idle compute units, mitigating load imbalance and enhancing overall performance.
    Our experiments demonstrate that \textit{Nexus Machine} achieves 90\% better performance compared to state-of-the-art (SOTA) reconfigurable architectures, within the same power budget and area. \textit{Nexus Machine} also achieves 70\% higher fabric utilization, in contrast to SOTA architectures.

\end{abstract}


\maketitle

\section{Introduction}
Edge devices require high performance within constrained power budgets. ASIC accelerators are efficient for specific applications, but reconfigurable computing offers a flexible alternative with lower engineering costs. A reconfigurable architecture consists of many processing elements (PEs) connected by an on-chip network. 
Field Programmable Gate Arrays (FPGAs) are the most successful reconfigurable architectures today; however they suffer from sub-optimal compute density and energy efficiency due to bit-level reconfigurability.
In contrast, modern reconfigurable architectures provide word-level reconfigurability, balancing adaptability and efficiency with significant research targeting both edge~\cite{adres, hycube, snafu, riptide} and server domains~\cite{fifer, plasticine, capstan, amber}.

\begin{figure}[t!]
	\scriptsize
	\centering
	\includegraphics[width=\columnwidth]{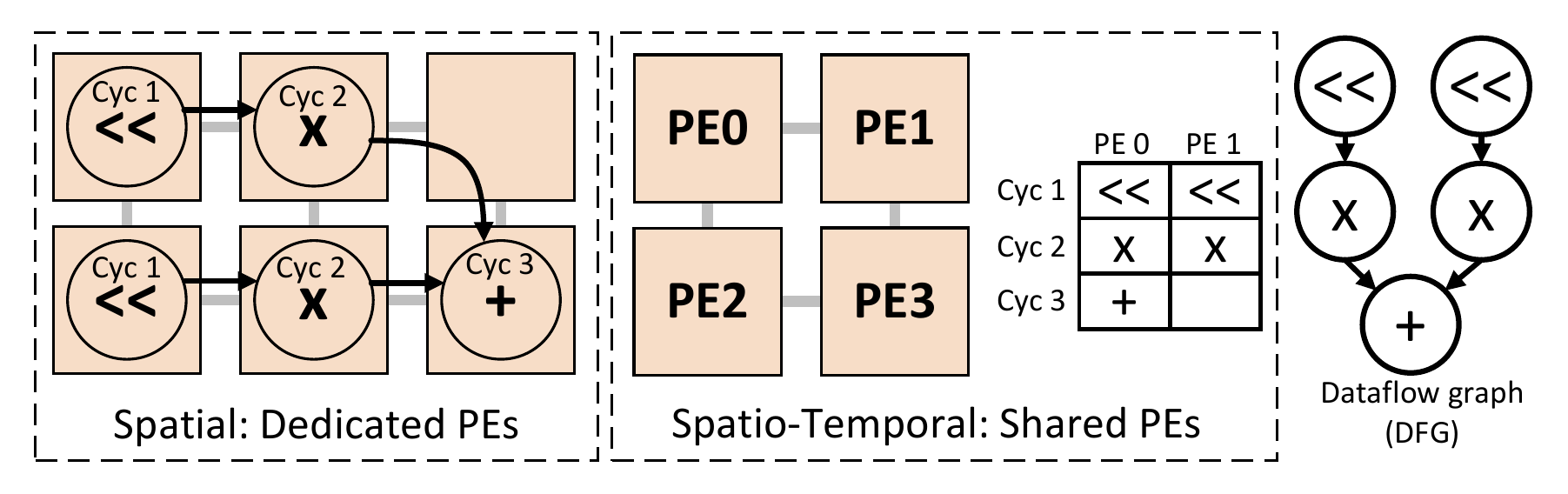}
	\caption{Spatial vs Spatio-temporal}
	\label{fig:taxanomy}
\end{figure}

A Coarse-Grained Reconfigurable Architecture (CGRA) uses simple PEs connected by a reconfigurable on-chip network to form high-throughput datapaths. Each PE contains an ALU, router, and reconfiguration memory. Data is loaded from a global scratchpad memory (SPM) to the PE array for execution, and results are stored back in the SPM. As shown in Fig.~\ref{fig:taxanomy}, CGRAs are classified into Spatial and Spatio-temporal~\cite{revel}. 
Spatial architectures~\cite{snafu, softbrain, piperench, riptide} maintain a fixed mapping of compute and communication, ensuring low configuration energy but potentially lower performance. 
Spatio-temporal~\cite{adres, hycube, trips, wavescalar}, on the other hand, allow each PE to reconfigure to a new instruction every cycle, enhancing performance but potentially increasing energy consumption~\cite{flex}. These architectures excel in accelerating regular workloads with predictable access patterns and minimal control flow, such as DSP workloads including dense linear algebra.

Critical workload domains for edge devices, such as edge AI, AR/VR, electric vehicles (EVs), gaming, and IoT, exhibit irregular computational kernels like sparse linear algebra and graph analytics.
These workloads become especially challenging with dynamic sparsity~\cite{pit}, such as activation and weight pruning in neural networks~\cite{teal_activationsparsity}, where both matrices are defined at runtime.
Current reconfigurable architectures struggle with irregular workloads, characterized by unpredictable memory accesses and complex control flow. These challenges arise because such dataflows are not well-suited for static compile-time instruction and data mapping techniques typically employed by these architectures. Irregular memory accesses lead to significant bank conflicts, limiting memory-level parallelism, while irregular control flow causes serious load imbalance, leaving most PEs idle at runtime. 

Recent works aim to tackle irregular workloads, but achieving high performance with energy efficiency remains challenging. Modern reconfigurable architectures like Triggered Instruction Architecture (TIA)~\cite{tia} leverage a data-local execution model for edge devices. Unlike CGRAs with centralized data memory, each PE in TIA has its own distributed memory and configuration memory. TIA operates by dispatching operands to a PE, triggering the loading of a statically placed instruction. These architectures rely on a runtime scheduler for tag matching and a priority encoder to trigger the next instruction, adding significant hardware overhead. Due to the irregular nature of workloads, these architectures face difficulties with load balancing and efficient fabric utilization.

\begin{figure}[h!]
	\scriptsize
	\centering
    \includegraphics[width=0.9\columnwidth]{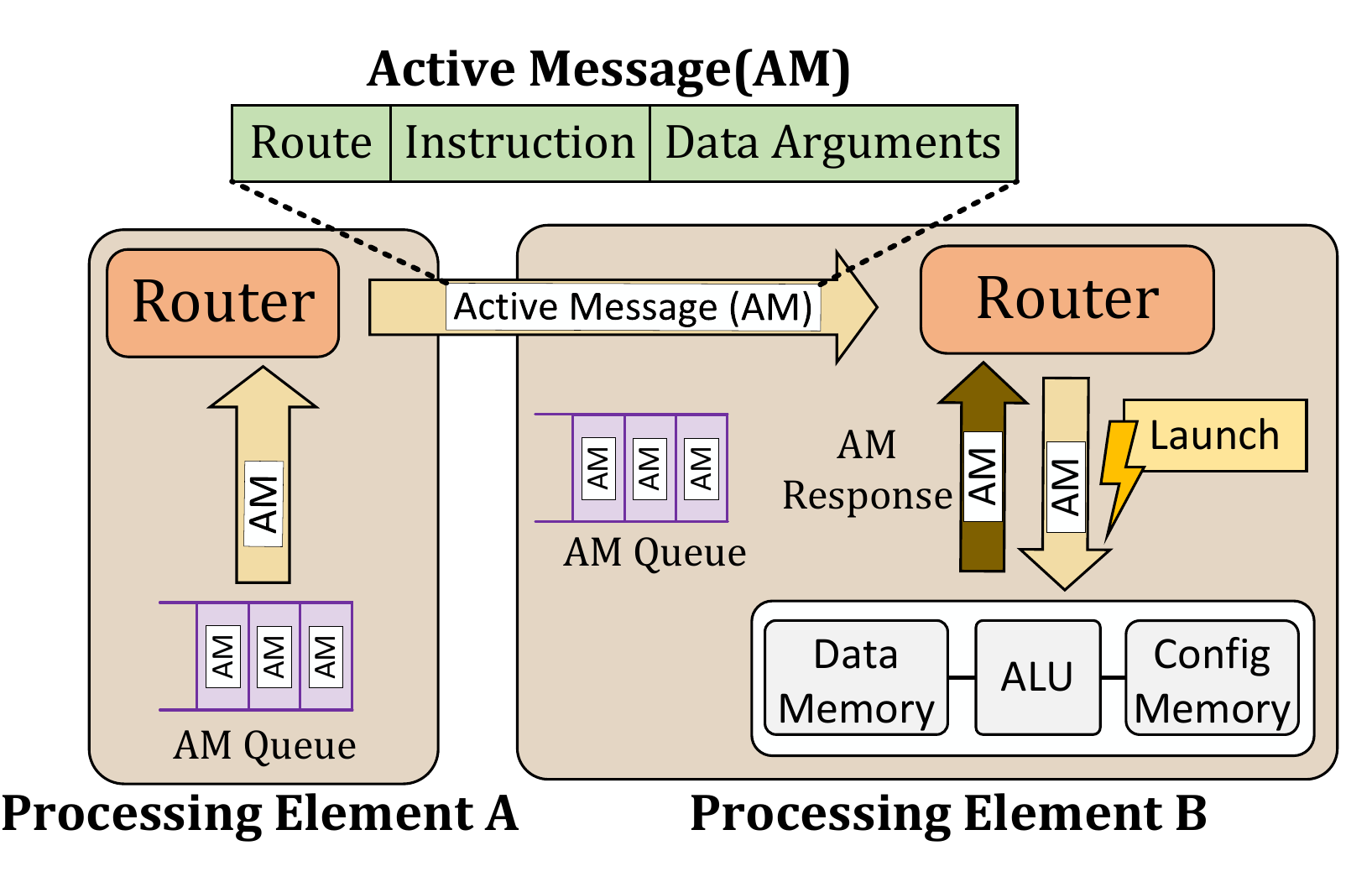}
	\caption{Active Message (AM) communication mechanism: An AM originating from PE A is launched at destination PE B, where it is executed on the ALU, and interacts with the data and configuration memory. Additional AMs may be generated in response, if necessary.}
	\label{fig:active_message}
\end{figure}
To address these challenges, we introduce \textit{Nexus Machine}, a novel reconfigurable architecture inspired by the Active Message paradigm. \textbf{Active Message (AM)}~\cite{am_culler} is a communication mechanism that enables flexible and efficient task execution by leveraging the proximity of tasks to relevant data. As shown in Fig.~\ref{fig:active_message}, each AM contains routing information, instructions for the target node, and data payload for arguments. An AM originating from PE A traverses the on-chip network to reach PE B, where it launches computation on the ALU, interacting with data and configuration memory. PE B can respond by generating additional AMs.

\textit{Nexus Machine} efficiently executes irregular workloads by leveraging the Active Message paradigm, addressing irregular memory accesses and load balancing. By distributing sparse tensors across the fabric and performing data-local execution, \textit{Nexus Machine} avoids expensive bank conflicts and enhances memory access. It provides a specialized compiler framework and hardware for \textit{Data-driven execution} of sparse tensors, optimizing memory accesses. To improve load distribution and system efficiency, \textit{Nexus Machine} introduces a novel load-balancing strategy for In-Network execution of AMs, dynamically deploying and executing AMs en-route.

Compared to contemporary platforms, while CPUs and GPUs are often favored for their flexibility and ease of programming, they are less efficient for highly irregular workloads compared to domain-specific architectures. GPUs perform well with minimal control divergence, but optimizing them for irregular workloads remains an active research area~\cite{irregular_gpu,irregular_gpu1}. Similarly, High-Level Synthesis (HLS) for FPGAs is optimized for regular patterns, limiting its effectiveness for irregular workloads, a key focus of ongoing research~\cite{irregular_fpga,irregular_fpga2}. Unlike FPGAs, which reconfigure at the gate level with a bitstream, \textit{Nexus Machine} reconfigures its ALUs instruction-by-instruction each cycle, offering faster reconfiguration times. Additionally, unlike some previous CGRAs~\cite{dynamic_ii} that use LUTs with ALUs within their PEs for finer control, \textit{Nexus Machine} does not use LUTs. Our contributions include:
\begin{itemize}
    \item An architecture and compiler framework for \textit{Nexus Machine}, showcasing its applicability in sparse tensor computations.
    \item An exhaustive exploration of \textit{Nexus Machine}'s architectural parameters through a cycle-accurate simulator, covering diverse workloads.
    \item Implementation of \textit{Nexus Machine} in 22nm process with compiled memories, outperforming a \textit{generic CGRA} baseline by 1.9x and achieving 1.7x better fabric utilization.
\end{itemize}
\vspace{-1mm}
\section{Background and Motivation}
\label{section:motivation}
\subsection{Active Messages and Task Migration}
Active Messages (AM)~\cite{am_culler} was proposed as a lightweight communication scheme that sends network messages from one machine to another, with messages triggering execution of user-defined functions upon arrival.
These lightweight functions can access memory, invoke computations, and dispatch further Active Messages as a response.
The core concept behind Active Message revolves around dispatching computation to the data location rather than transferring data to the computation.
This effectively removes coherency overhead and minimizes data movement.

AMs have been realized and implemented in a range of multi-computers in the past. In~\cite{am_culler}, a practical implementation of AM on CM5 and n-CUBE/2 architectures was presented.
The J-Machine~\cite{jmachine} is a notable parallel processor featuring comprehensive hardware and instruction support for active messages.
It facilitates the remote spawning of fine-grained tasks at a granularity of approximately 20 instructions.
Also, commercially available processors like the Cray T3D and its subsequent iterations have leveraged network designs similar to that of the J-Machine.
\begin{figure*}[h!]
	\scriptsize
	\centering
	\includegraphics[width=0.9\textwidth]{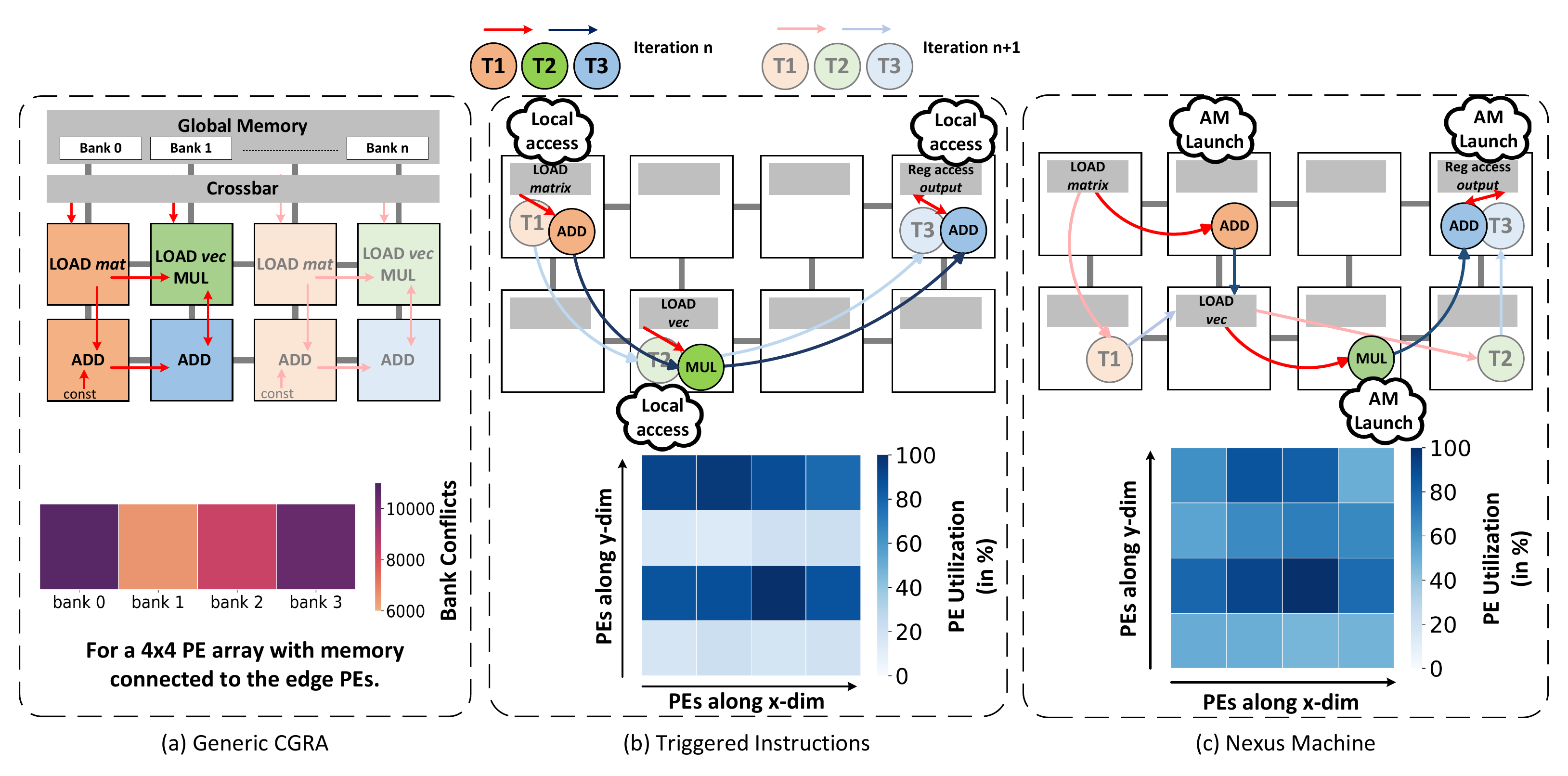}
    \vspace{-0.25cm}
	\caption{Program execution comparison for SpMV kernel, illustrating two consecutive iterations with a bank conflict. (a) \textit{Generic CGRA}: Data flows through statically placed instructions (top) and bank conflicts across various banks for a real workload with n=2048 on a 4x4 PE array (bottom) (b) Triggered Instructions: Illustrates data-local execution with messages, invoking tasks at the location of data, reducing data movement (top) and visual representation of the load imbalance across the PE array (bottom) (c) \textit{Nexus Machine}: Enhances performance and PE utilization through a unique approach — enabling opportunistic execution and utilizing idle ALUs for en-route instruction execution (top) and visual representation of the uniform load balance across the PE array (bottom). Data movements are represented by {\color{red} red} arrows, while {\color{blue} blue} arrows depict message transfers.} 
    \label{fig:walkthrough}
    \vspace{-.3cm}
\end{figure*}
\begin{figure}[h!]
	\scriptsize
	\centering	\includegraphics[width=0.9\columnwidth]{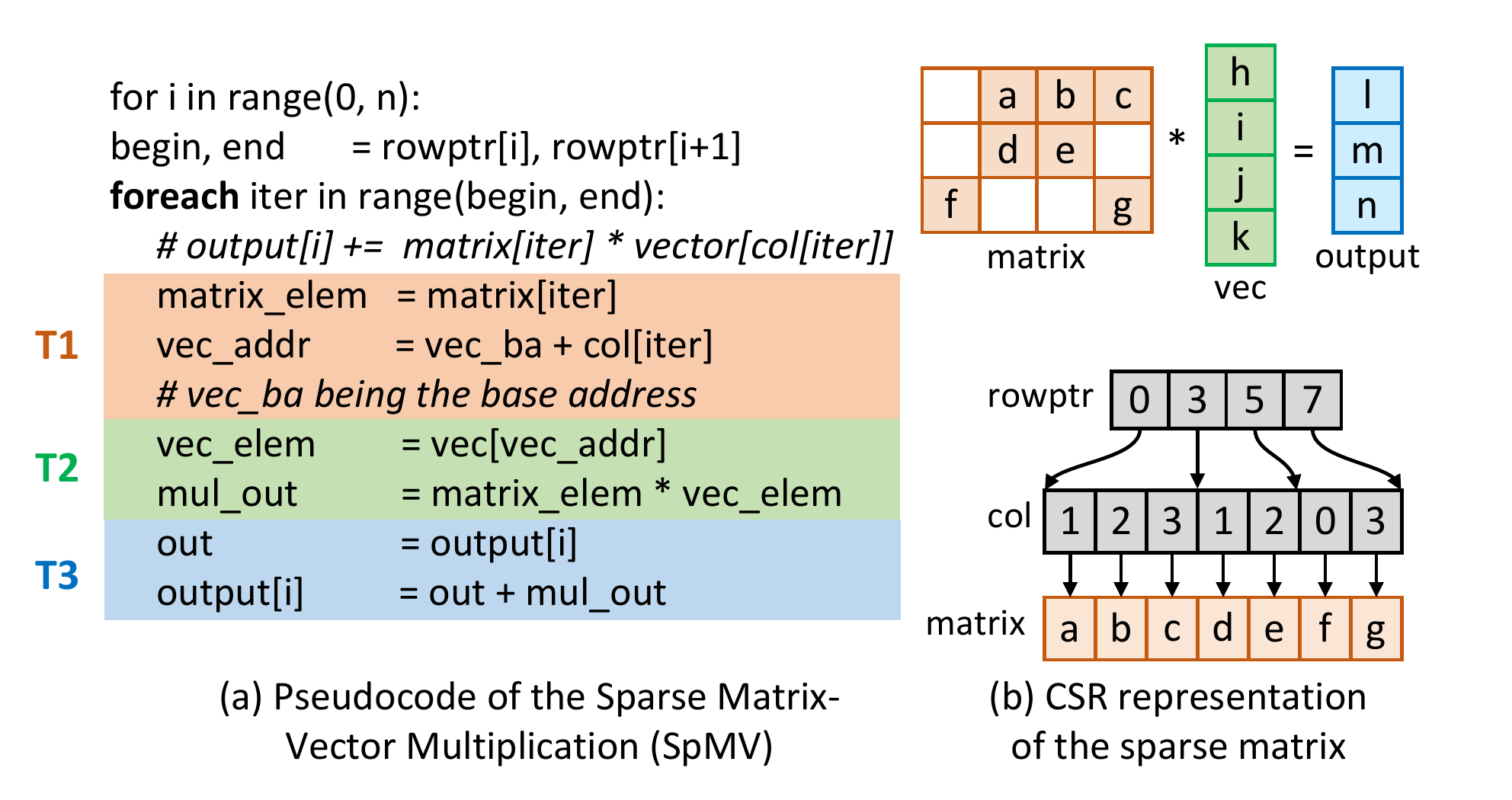}
	\caption{Sparse Matrix-Vector Multiplication (SpMV).} 
	\label{fig:motivation}
\end{figure}
With \textit{Nexus Machine}, we propose a low-power reconfigurable architecture that is inspired by the Active Message model to target irregular workloads. 
Our definition of AM diverges significantly from the original concept in its application to reconfigurable architectures. 
First, we store the instruction itself within the message, rather than relying on a handler to trigger a set of instructions, to ensure immediate execution of the instruction upon arrival at a PE. 
\textit{Nexus machine} also leverages the static compilation strategy of previous reconfigurable architectures, with the software compiler first intelligently placing instructions and data across the fabric.
Unlike previous architectures, where the data flows through statically placed instructions on PEs, \textit{Nexus Machine} reduces data movement by sending and executing instructions at the location where the data resides.

In the \textit{Nexus Machine} architecture, the entire execution is initiated and orchestrated by the data; thus we refer to it as \textbf{Data-Driven Execution}.
Unlike prior multi-computer systems with high bandwidth networks between machines, \textit{Nexus Machine} employs lightweight Networks on Chip (NoCs) between simple ALUs, similar to previous reconfigurable architectures, which facilitates frequent message passing to initiate ALU operations, making the \textit{Nexus Machine} well-suited for the \textit{Data-Driven Execution} model.

\subsection{Challenges of Irregular Workloads on reconfigurable architectures}
\label{section:nexusmachine_motivation}
\noindent

Current reconfigurable architectures excel in creating static software pipelining, where a task is divided into instruction-level pipeline stages, and gets mapped to the PEs across the fabric.
To execute tasks, data moves from one PE (producer) to the next PEs (consumers, data-dependent on the producer) along fixed (compile-time) pre-scheduled datapaths. 
This works well for regular workloads where the data access and control flow are predictable.

However, mapping irregular workloads on these architectures significantly impacts performance, fabric utilization, and energy consumption for data movement.
To illustrate this issue with an example, we use \textbf{Sparse Matrix Vector Multiplication (SpMV)} kernel that multiplies a sparse matrix \textit{matrix} with a dense vector \textit{vec}. 
Fig.~\ref{fig:motivation}(a) provides a pseudo-code for SpMV at the instruction level. The sparse matrix in Fig.~\ref{fig:motivation}(b) is stored in a commonly used compressed sparse row (CSR) format to encode the row and column indices of the non-zero elements. \textit{rowptr} points to the first non-zero element information for each row in the \textit{col} vector, which in turn stores the column indices of all the non-zero elements for each row consecutively.
SpMV is an irregular workload due to its use of multiple levels of indirection to access the data: \textit{rowptr} to access \textit{col}, which in turn helps access a specific index in \textit{vec}.

Fig.~\ref{fig:walkthrough}(a) (top) shows the mapping of SpMV on a \textit{generic CGRA} that has a global memory shared across all the PEs. Here the operations are statically mapped onto the PEs and the routing path of the data between dependent PEs is also fixed (red arrows).
Further, to enhance fabric utilization, multiple iterations of the loop are unrolled and spatially mapped.
However, this approach exacerbates bank conflicts when dealing with irregular workloads. 
The architecture's demand for synchronized operation across all PEs in a predictable manner means that any bank conflict results in stalls.
For the SpMV workload mapped to an array of 16 PEs connected to 4 memory banks on one edge, Fig.~\ref{fig:walkthrough}(a) (bottom) illustrates the occurrences of conflicts across different memory banks.

To mitigate bank conflicts caused by irregular memory access patterns in global memory, we distribute the memory across PEs, enabling data-local execution that serializes memory accesses at each PE. 
To convert SpMV into a sequence of tasks suitable for data-local execution, we divide the code at each memory operation. This is illustrated for SpMV by the different colored segments in Fig.~\ref{fig:motivation}(a).
Task \textbf{T1} accesses the \textit{matrix} array (represented by \textit{rowptr} and \textit{col}), \textbf{T2} accesses elements in the \textit{vec} array, and \textbf{T3} executed on the corresponding elements of the \textit{output} array.
During the compilation phase, \textit{Nexus Machine} restructures the CSR arrays of the matrix into AM entries (see Section ~\ref{section:compiler}). 
Each entry consolidates the matrix data and the locations of vector and output elements, using \textit{rowptr} and \textit{col} arrays. 
These entries help guide the messages as they traverse the architecture.
In \textit{Nexus Machine}, dynamic routing introduces network congestion analogous to bank conflicts in global memory CGRAs, as message routing depends on congestion levels. 

The architecture baseline we use for data-local execution resembles that of Triggered Instructions~\cite{tia} with multiple PEs, where each PE includes a distributed data memory, ALU, and a configuration memory with a scheduler. The tasks are anchored with the data, triggered when messages arrive with the data. 
Fig.~\ref{fig:walkthrough}(b) (top) illustrates how our transformed SpMV can be executed on such an architecture.
\textbf{T1} performs local load for the \textit{matrix} element and calculates the address for the \textit{vec} element.
When the message reaches the PE with the \textit{vec} element, \textbf{T2} executes a local load of \textit{vec} element and multiplies it with the \textit{matrix} element obtained from the remote PE.
Finally, when the message reaches the PE with the \textit{output} element, \textbf{T3} performs a local aggregation.
While our transformed SpMV workload can be mapped to such an architecture, it still faces challenges in terms of performance and resource utilization due to load imbalance. This is because the instructions are anchored with the data, and cannot move at run time to different PEs.
For SpMV, these imbalances are evident by the varying numbers of elements aggregated for each \textit{output}.
Fig.~\ref{fig:walkthrough}(b) (bottom)  represents the load imbalance observed within PE array.

\textit{Nexus Machine} effectively addresses the low utilization and performance challenges posed by irregular workloads by opportunistically executing active messages on the idle PEs en-route.
It does this by exploiting the AM paradigm that allows the transmission of not just data operands, but also {\bf instructions} within a single message.
The architecture employs \textbf{Opportunistic Execution}, a strategy that postpones execution along the datapath instead of executing immediately on the source PE. Operations are executed on the first idle PE encountered along the route, or alternatively, on the destination PE if no idle PEs are available.
In Fig.~\ref{fig:walkthrough}(c) (top), we illustrate how \textit{Nexus Machine} applies this approach to execute the SpMV kernel.
Task \textbf{T1} initiates the local access of the \textit{matrix} element but forwards the computation into the network, where it is executed on the first available idle PE. 
This generates an AM response with Task T2.
Similarly, task \textbf{T2} is sent into the network with the corresponding \textit{matrix} and \textit{vec} elements and is computed en-route, resulting in another AM response with Task T3.
Finally, task \textbf{T3} performs a local aggregation of the \textit{output} element. 
We refer to this as \textbf{In-Network Computing} for dataflow architectures.
Fig.~\ref{fig:walkthrough}(c) (bottom) visually demonstrates how \textit{Nexus Machine}'s coupling of compiler data placement with runtime distribution of instructions across the fabric enables uniform load balance across the PEs, significantly enhancing utilization and performance.
\section{Nexus Machine}
\textit{Nexus Machine} is a novel reconfigurable architecture specialized for irregular workloads that uses AM paradigm for \textit{Data-Driven} execution model. This section introduces the execution model of \textit{Nexus Machine}, followed by its micro-architecture and compiler.
Fig.~\ref{fig:detail_arch}(a) provides an overview of the architecture, showcasing PEs interconnected via a mesh network.
\subsection{Execution Model}
\textbf{Static Initialization.}
\textit{Nexus Machine} follows a distributed tensor placement approach, partitioning all tensors across the PEs.
Initially, the compiler generates a \textit{static AM} for each element of the first tensor.
These \textit{static AMs} are then stored in the active message network interface, while the remaining tensors are placed in the data memory within each PE. 
Additionally, the compiler generates opcodes corresponding to the workload and stores them in the configuration memories of all the PEs.

\textbf{Dynamic Execution.}
\begin{figure}[t!]
	\scriptsize
	\centering
    \includegraphics[width=0.88\columnwidth]{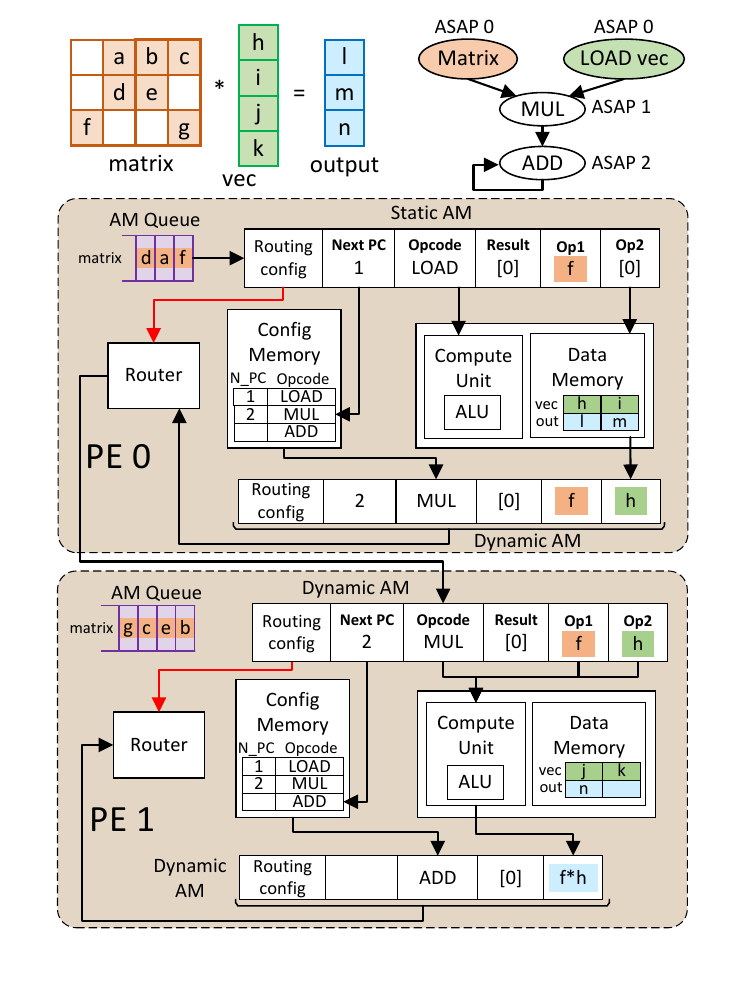}
    \vspace{-0.65cm}
    \caption{Execution of SpMV using the data in Fig.~\ref{fig:motivation} on a fabric with 2 PEs. It illustrates the placement of matrix, vector, and output partitions, along with AM generation. [] denotes element address, and {\color{red}red} arrows represent control signals.}
    \label{fig:exec_model}
\end{figure}
When execution is initiated, the active message network interface dequeues the first \textit{static AM} and then routes it based on dynamic turn model~\cite{noc_peh} routing protocol to the PE containing the next operands.
Upon arrival at the PE, the AM is decoded and updated accordingly, transforming it into a \textit{dynamic AM} created on-the-fly, in contrast to the predefined \textit{static AMs} generated at compile time.
Once the AM has gathered all the required operands, it proceeds towards the next destination PE, for execution. 
Additionally, we allow these AMs to perform computations en-route if they encounter an available compute unit.
The final result gets stored in the memory of the destination PE.

Thus, \textbf{static AMs} are generated at compile time, containing the initial instructions for computation. During execution, \textit{static AMs} are transformed into \textbf{dynamic AMs}, carrying different instructions based on the AM format and real-time conditions. This allows \textit{Nexus Machine} to optimize both data placement and computation, enhancing efficiency for irregular workloads.

Fig.~\ref{fig:exec_model} illustrates the execution of SpMV using the example from Fig.~\ref{fig:motivation} on a fabric with 2 PEs. The \textit{matrix} operand is split and converted into \textit{static AMs}, stored in \textit{AM Queues}, while \textit{vec} and \textit{output} are divided and placed in \textit{data memories}. At cycle 0, PE0’s initial \textit{static AM} (with operand f) performs a LOAD to fetch \textit{Op2} (h in this case) from \textit{data memory}, then creates a \textit{dynamic AM} with \textit{Op2} and \textit{Opcode} MUL. This \textit{dynamic AM} is sent to PE1, where operands are multiplied in cycle 1, and the result is combined with \textit{Opcode} ADD to generate the next AM. PE1 updates \textit{output} (n in the \textit{data memory}) by adding the multiplied result to it (not shown in the figure).
Note that PE1 concurrently dequeues and processes AMs from its \textit{AM Queue}, although this is not shown here for the sake of simplicity.

\begin{figure}[t!]
	\scriptsize
	\centering	\includegraphics[width=0.85\columnwidth]{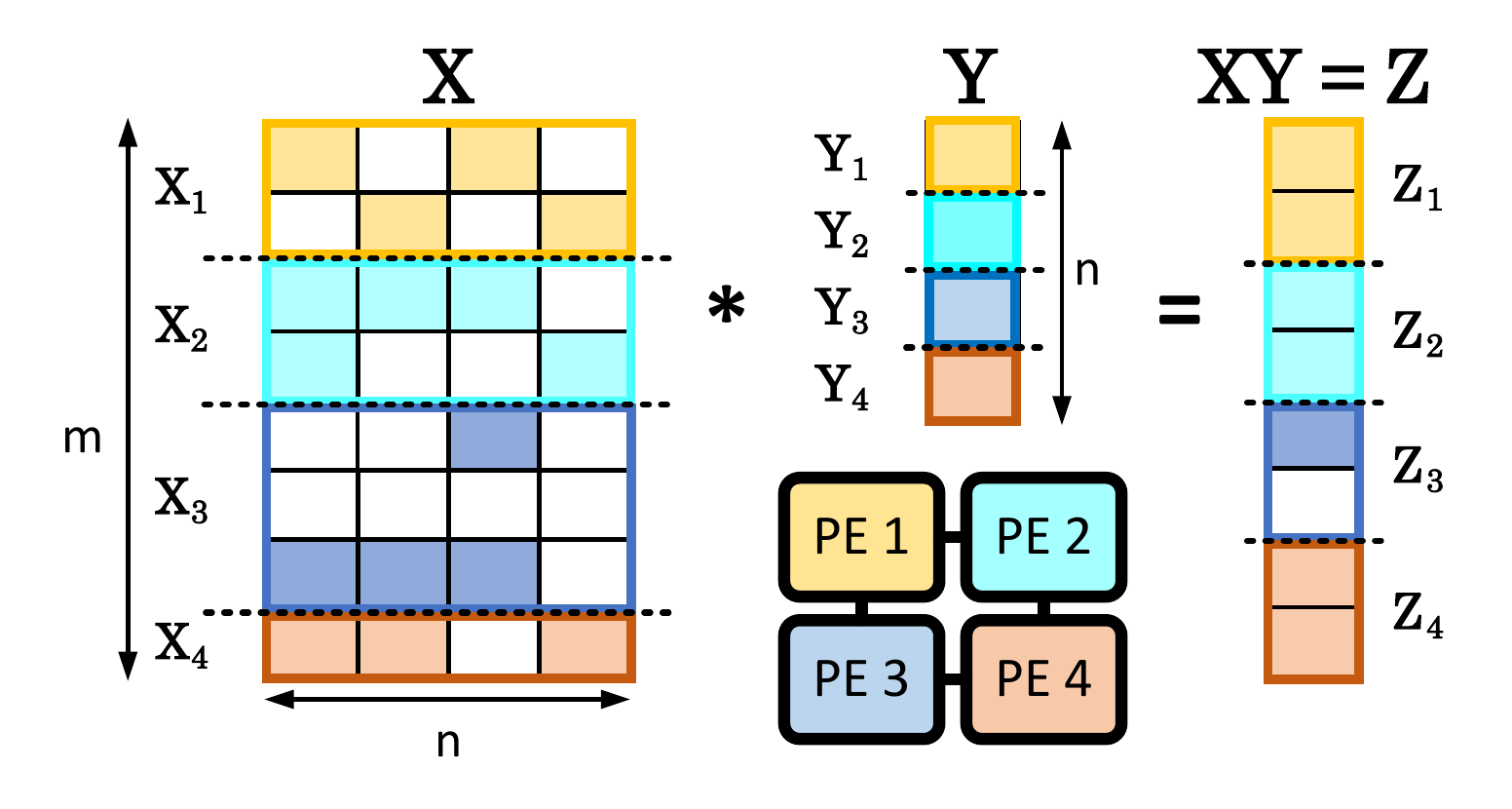}
    \vspace{-0.25cm}
    \caption{Data partitioning for SpMV involves multiplying a sparse matrix \textbf{X} ($m \times n$) with a dense vector \textbf{Y} ($n \times 1$) to produce \textbf{Z} ($m \times 1$). Four partitions are shown in different colors, distributed across PEs denoted by the same color. Non-zero elements are shown in colored squares.}
	\label{fig:data_partitioning}
\end{figure}

{
\subsubsection{Tensor Partitioning.}
\label{section:data_placement}
\textit{Nexus Machine} employs a coarse-grained distributed tensor partitioning strategy using a dissimilarity-aware approach. Each tensor is segmented across PEs to balance memory utilization and reduce communication overhead. For large tensors exceeding local capacity, tiling~\cite{extensor, tiling} decomposes the computation into smaller sub-tensors that fit within on-chip memory.

Fig.~\ref{fig:data_partitioning} illustrates an SpMV example for $\textbf{N}=4$ PEs. The sparse input tensor $\textbf{X}$, stored in Compressed Sparse Row (CSR) format, is divided into $\textbf{X}_1, \dots, \textbf{X}_N$, each with the same column dimension but differing row counts, such that $nnz(\textbf{X}_i) \approx nnz(\textbf{X})/N$. This partitioning is computed via a linear scan of the row pointer array, with complexity $\mathcal{O}(m)$, where $m$ is the number of rows. The one dimensional tensors $\textbf{Y}$ and $\textbf{Z}$ are partitioned correspondingly to maintain alignment with $\textbf{X}$. For dense tensors, uniform segmentation into $k$ equal parts is applied. As described in Sec.~\ref{section:compiler}, partitioning plays a critical role in balancing locality and execution efficiency, and is tightly integrated into the compilation flow to reflect sparsity patterns and architectural constraints.
}

\subsubsection{Data-Driven Execution.}
In the \textit{Nexus Machine} architecture, each PE stores the first input tensor as pre-compiled active messages (AM) and a partition of other tensors. At the source PE, the first AM is sent to the PE with the next data element. Since the initiation of execution is triggered by the active message or one of the tensor elements, we refer to this model as \textit{Data-Driven Execution}. 

Fig.~\ref{fig:detail_arch}(a) illustrates an example where instruction \textit{I0} at PE4 triggers \textit{I1}, generating a dynamic AM with \textit{I1} and \textit{I0}'s output, then sent to destination PE15 for execution.
However, restricting execution solely to the source or destination PE results in significant underutilization and workload imbalance across PEs for irregular workloads where data dependencies and control flow are unpredictable.

\subsubsection{In-Network Computing.}
To enhance load balancing, we employ \textit{opportunistic execution} approach, allowing an AM to execute on an \textbf{intermediate PE} as it travels towards its final destination.
With this execution model, en-route AMs carry both the instruction and required data operands, enabling \textit{intermediate PEs} to perform computations whenever an idle ALU is encountered.
In Fig.~\ref{fig:detail_arch}(a), the highlighted cores in blue represent the potential \textit{intermediate PEs} for a message routed from PE4 to PE15.
The message is executed at PE13, denoted as \textit{I1'}, as it serves as the first intermediate PE along the route with an idle ALU.

This provides several advantages: 
(a) It introduces a hardware mechanism to enhance workload distribution and fabric utilization by leveraging idle PEs,
(b) reduces NoC contention by deciding whether messages are executed en-route or continue toward their destination PEs, and
(c) minimizes the amount of data traversing the NoC by coalescing the updates to the original message and discarding unnecessary data.

\subsubsection{Termination and Global Synchronization.}
\textit{Nexus Machine} completes execution when all PEs are inactive and no messages are in transit, generating a global idle signal to indicate completion. \textit{Nexus Machine} uses this global idle signal to notify the host via an interrupt.  This approach is suitable for edge architectures with limited memory, where tiling improves resource use and scalability. Data tiles are executed sequentially in a global synchronized manner. Once a tile finishes, the system detects the idle state and triggers the next tile's execution.
\subsection{Multi-destination active message format}
\label{section:message_format}
\vspace{-0.7cm}
\begin{figure}[h!]
	\scriptsize
        \centering
    \includegraphics[width=1\columnwidth]{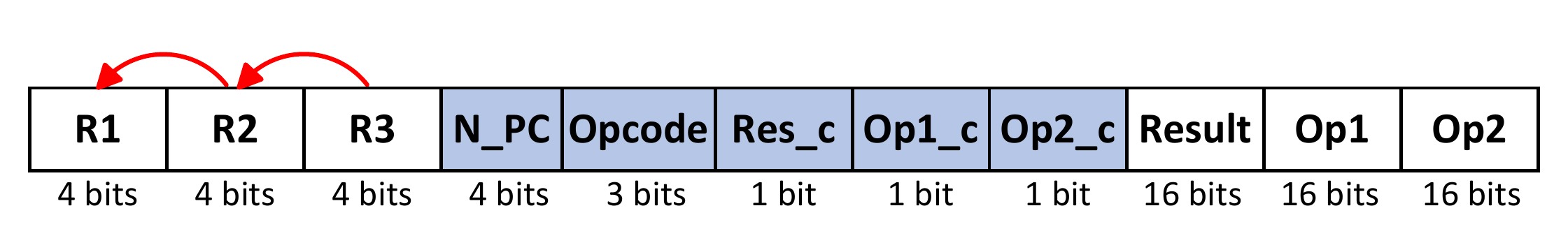}
	\caption{Active Message format} 
	\label{fig:message_format}
\end{figure}
\begin{figure*}[t!]
	\scriptsize
	\centering
	\includegraphics[width=0.9\textwidth]{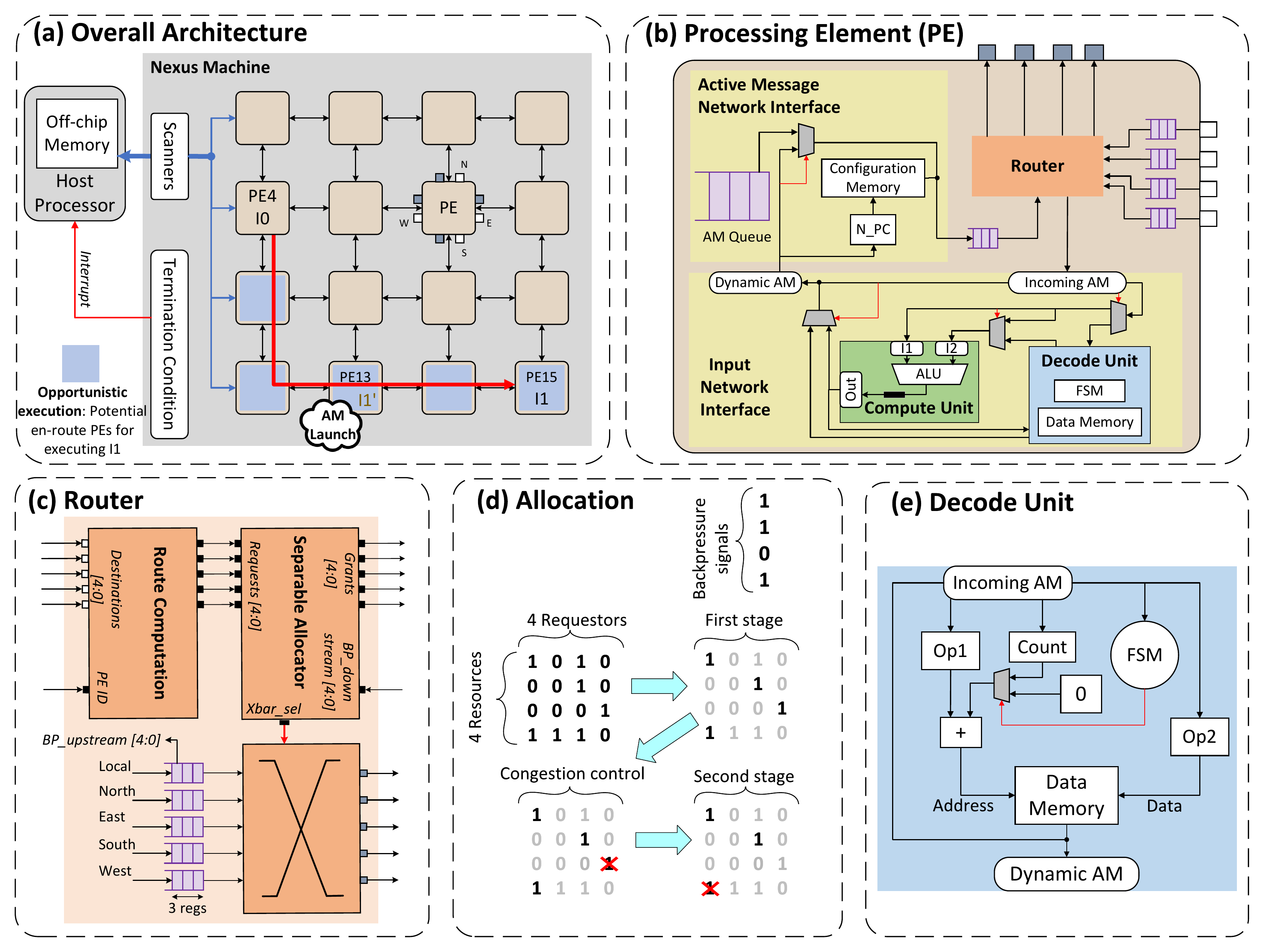}
	\caption{\textit{Nexus Machine} microarchitecture. A fabric of homogenous PEs interconnected by a mesh network for communicating Active Messages which carry instructions that can be launched en-route at any PE, enhancing fabric utilization and runtime.} 
    \vspace{-0.5cm}
	\label{fig:detail_arch}
\end{figure*}
\textit{Nexus Machine} extends the fundamental Active Message primitives to accommodate a multi-destination based routing mechanism. 
Fig.~\ref{fig:message_format} illustrates the message format: the first 12 bits specify intermediate destinations (\textit{R1}, \textit{R2}, \textit{R3}), based on our workload analysis (as SDDMM has three inputs, destinations correspond to two inputs and one output tensor). 
The next 4 bits contain the Program Counter (PC) for the next instruction (\textit{N\_PC}), followed by 3 bits for the \textit{Opcode}.
A single bit (\textit{Res\_c}) indicates if the message carries a result. 
The subsequent 2 bits (\textit{Op1\_c} and \textit{Op2\_c}) identify whether \textit{Op1} and \textit{Op2} are addresses or values. 
Depending on \textit{Res\_c}, the \textit{Result} field contains the final result or its address, while the next 16 bits hold data for Operand1 (\textit{Op1}) and Operand2 (\textit{Op2}).

{When the message reaches its first destination (\textit{R1}), it is routed through the LOCAL port of the corresponding PE and forwarded to its ALU for execution. The ALU processes the associated computation payload, after which the message is reassembled with an updated destination list. The remaining destinations are cyclically rotated, making \textit{R2} the first and \textit{R3} the second.
} In the \textit{Nexus Machine}, a message is equivalent to a packet or flit (all messages are a single-flit packet).
\subsection{Nexus Machine Micro-architecture}
As presented in Fig.~\ref{fig:detail_arch}(a), the \textit{Nexus Machine}'s fabric comprises processing elements (PEs) interconnected with a mesh network, with a global termination detector. Each PE is linked to four neighboring PEs in North, East, South, and West directions.
The off-chip memory is connected to the four PEs located along the left edge.

\subsubsection{Processing Elements (PEs).}
As shown in Fig.~\ref{fig:detail_arch}(b), each PE has a compute unit, a dynamic router for network connectivity with congestion control, a decode unit, and two Network Interface logic.
\textit{Input Network Interface} unit is responsible for efficiently handling incoming AMs from the NoC, while the AM Network Interface unit initiates the injection of new messages into the NoC.

\textbf{Input Network Interface.}
The \textit{Input Network Interface} unit manages \textit{incoming AMs} to a PE. Depending on the message:
(a) If it pertains to an ALU operation, it is directed to the \textit{Compute Unit} for execution; (b) Alternatively, in case of a memory operation, the message is forwarded to the \textit{Decode} unit. 
This unit initiates a load or store operation, utilizing the operand address information (\textit{Op1} or \textit{Op2}) contained in the message.
Once these operations are completed, the resulting \textit{output dynamic AM} is dispatched to the \textit{AM Network Interface} for injecting into the network.

\textbf{Compute Unit.}
The \textit{compute unit} within a PE can perform 16-bit arithmetic operations, logic operations, multiplication, and division on its ALU. An incoming AM at the \textit{Input Network Interface} dispatches two operands, \textit{Op1} and \textit{Op2} along with the \textit{Opcode} field in the message to the compute unit.
After computation, it generates an output that is combined with the original AM, replacing the \textit{Op1} field in the message.
Finally, this modified AM is forwarded to the \textit{AM Network Interface} for injecting into the network.


\textbf{Decode Unit.}
The \textit{Decode Unit}, as shown in Fig.~\ref{fig:detail_arch}(e), can be flexibly configured to operate in dereference and streaming modes.
In \textbf{dereference mode}, the operand address field (\textit{Op1} or \textit{Op2}) in the message triggers the loading of a single element. This gets embedded into the output \textit{dynamic AM}.
Conversely, in \textbf{streaming mode}, the message initiates the loading of multiple elements from memory, generating multiple output AMs.
In this mode, the operand address is considered the base address, along with a count to access and load the elements from memory sequentially.
These two modes suffice for our benchmarks; however, our architecture allows for integration of additional modes if needed.

\textbf{Active Message (AM) Network Interface.}
The \textit{AM Network Interface logic} is responsible for injecting AMs into the network.
This module comprises two primary components: an \textit{AM Queue} and a \textit{configuration memory}. 
{Each PE contains a dedicated 1KB \textit{AM Queue} (16KB total across the 16-PE array), implemented as a FIFO that stores 70-bit precompiled entries. The \textit{configuration memory}, 10 bits wide, supports up to 8 distinct configurations per PE.}

As illustrated in Fig.~\ref{fig:detail_arch}(b), this module performs one of these two operations:
(1) If the output \textit{dynamic AM} is available from \textit{Input Network Interface}, the subsequent configuration is loaded from memory based on the \textit{N\_PC} field of the AM (see Fig.~\ref{fig:message_format}). 
This configuration is combined with the output \textit{dynamic AM} and forwarded into the injection port of the router.
(2) Alternatively, a \textit{static AM} is injected into the network to keep it occupied. 
This \textit{static AM} is the concatenation of the next precompiled entry from the \textit{AM Queue} with the first configuration loaded from memory.
The generation rate of \textit{static AMs} is determined by the backpressure signal at the router's injection port. The highlighted blue fields in the message format (see Fig.~\ref{fig:message_format}) depict data from the configuration memory used to construct the subsequent dynamic AM, with fields \textit{Res\_c}, \textit{Op1\_c}, and \textit{Op2\_c} stored to prevent redundancy. 



\subsubsection{Dynamic and Congestion Aware Routing.}
\textit{Nexus Machine} supports turn model routing~\cite{noc_peh}, with each router containing five input and five output ports.
Specifically, these input ports are designated for messages coming from \textit{AM Network Interface} unit, as well as north, east, south, and west directions, whereas output ports are designated for messages going to \textit{Input Network Interface} unit and four directions.
Each input port has a buffer comprising three registers to manage in-flight messages, accompanied by congestion control logic. \textit{Nexus Machine}'s design choice of employing only three registers is motivated by the goal of minimizing overall power consumption.
As presented in Fig.~\ref{fig:detail_arch}(c), each router contains a Route Computation Unit, Separable Allocator, and a Crossbar.

\textbf{Route Computation} logic considers the destination of messages from all the input ports. It compares it with the positional ID of the PE, and calculates the output port to be requested. This is sent as an input to the allocator. A toy example of \textbf{Separable Allocation} process is presented in Fig.~\ref{fig:detail_arch}(d)~\cite{noc_peh}.

\textbf{On/Off Congestion control} involves the transmission of a signal to the upstream router when the count of available buffers falls below a threshold, ensuring all in-flight messages will have buffers on arrival. Each of the five ports transmits an OFF signal when their corresponding available buffer space is reduced to 1, i.e., $T_{OFF} = 1$, and an ON signal when their buffer space reaches 2, i.e., $T_{ON} = 2$. The allocator output controls the 6x5 \textbf{Crossbar}.

\subsubsection{Off-chip Memory Datapath.}
Each off-chip memory port connects to a row of the PE array via an AXI bus, delivering a combined bandwidth of 1.28GBps. During data loading, data transfers from off-chip memory to the \textit{AM queues} and \textit{data memory} in each PE. 
The \textit{AM queues} are actively consumed during execution, effectively hiding data loading latency by performing it concurrently with the execution. 
However, data loading into \textit{data memories} occurs after each tile execution is complete. 

\subsubsection{Sparse Metadata Scanners.}
The first sparse operand is encoded in \textit{static AMs}. For subsequent sparse operands, a bit-vector scanner hardware assists in efficient iteration over sparse data, providing coordinates within compressed vectors as described in \cite{capstan}. \textit{Nexus Machine} integrates a modified version of the scanner with its AXI bus controller to obtain these coordinates. It can handle decoding of vectors of 16 non-zeros and more within 128 elements, allowing it to handle matrices with densities exceeding 12\%.
\vspace{-0.2cm}
\subsection{Deadlock avoidance}

Given the dynamic nature, \textit{Nexus Machine} can potentially encounter deadlocks without careful design. We address various deadlock scenarios with specific design choices: 
(1) To mitigate flow control deadlocks within the network, we adopt the bubble NoC~\cite{bubble_flow} approach over Virtual Channels (VCs), with the aim to minimize buffering. 
(2) Routing deadlocks are mitigated by using the turn model~\cite{noc_peh}, that ensures high throughput without complex adaptive routing hardware.
(3) AMs can potentially create deadlocks between the network and PEs. 
These are mitigated by the compiler through strategic data placement and runtime timeouts.
Future research will explore more optimized data placement strategies.
\subsection{Programming \textit{Nexus Machine}}
{\textit{Nexus Machine} targets affine loops with associative operations, using the polyhedral model to enable parallel execution, especially for AI workloads involving reduction loops. Programs are written in C, where the programmer annotates independent loop iterations with \textit{parallel-for} keyword, similar to OpenMP or CUDA. The compiler supports nested loops and fully unrolls annotated loops, maximizing parallelism across all levels. Each iteration is flattened, ensuring sequential execution of instructions within each iteration to maintain intra-iteration dependencies. It is the programmer’s responsibility to ensure correct annotations for non-associative inter-iteration dependencies. \textit{Nexus Machine} compiles and executes the program, running each \textit{parallel-for} loop iteration independently while preserving sequential semantics. Incorrectly annotating dependent loops as \textit{parallel-for} will result in undefined behavior. An example code illustrating these annotations is provided in Fig.~\ref{fig:exec_model}.

\subsection{\textit{Nexus Machine} Compiler}
\label{section:compiler}
\begin{figure}[h!]
	\scriptsize
	\centering
	\includegraphics[width=\columnwidth]{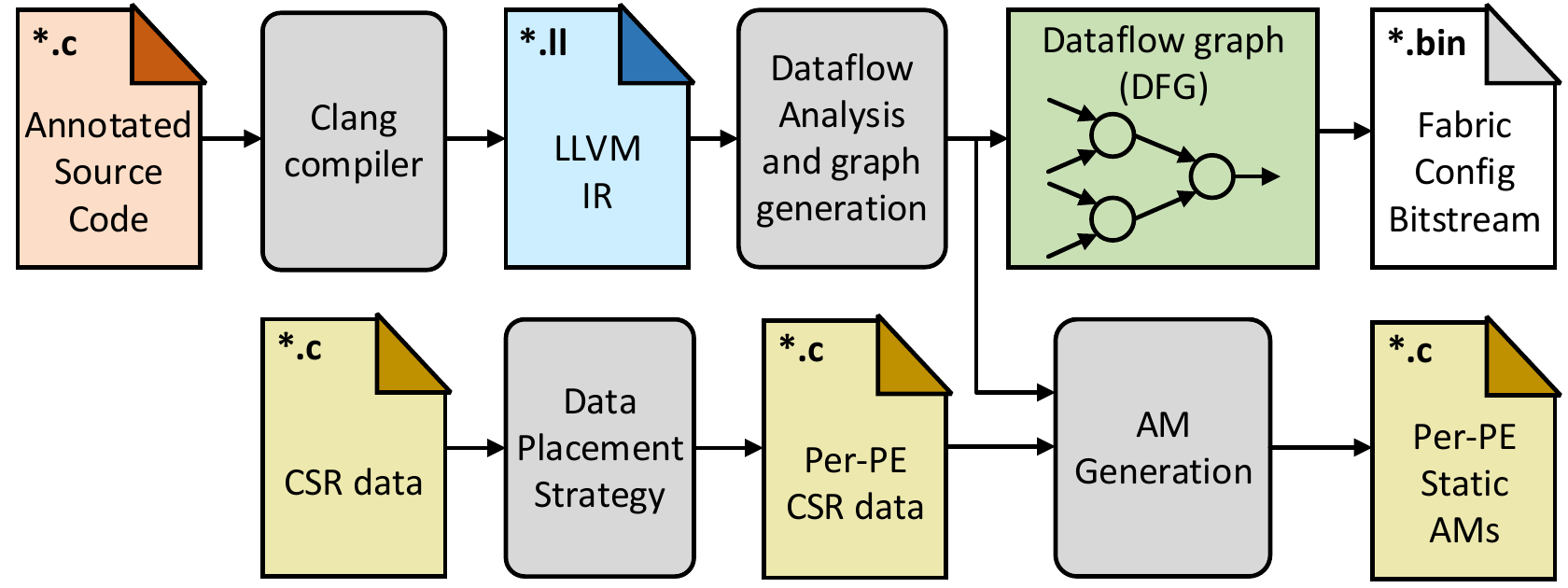}
	\caption{Process for transforming an application code and data into configuration and data format for \textit{Nexus Machine}.}
	\label{fig:compiler}
    \vspace{-0.5cm}
\end{figure}

\textit{Nexus Machine} integrates a static compiler responsible for preprocessing the application code, alongside a lightweight runtime manager executing on the host processor. 
The static compiler handles DFG (dataflow graph) generation, data partitioning, and data allocation. 
Concurrently, the runtime manager processes the data partitions, generating \textit{static AMs} that include the intermediate, final destinations and their designated locations for preloading into the \textit{AM Queues} within the PEs.
The dynamic routing is inherently done by \textit{Nexus Machine}'s architecture.

{\textbf{Static Compiler.} Fig.\ref{fig:compiler} illustrates the compilation flow for mapping annotated C code and data onto the \textit{Nexus Machine}. The front-end produces LLVM IR, from which a custom pass builds a DFG by identifying instruction dependencies and backedges. The DFG is scheduled using ASAP ordering, and the resulting opcodes are stored in configuration memory. Fig.\ref{fig:exec_model} shows a sample SpMV DFG.

\textit{Problem Definition for Distributed Data Placement:}  
Effective data placement is critical in this architecture, as it directly impacts performance and resource utilization. A fundamental tradeoff arises: \textbf{minimizing communication improves locality but limits possibilities for opportunistic AM execution; increasing communication enables idle PEs to assist via AMs, improving throughput.} Communication is significantly expensive due to network contention and energy constraints, making its reduction a key optimization target.

Given a set of sparse tensors \( A \) and a reconfigurable architecture with \( N \) homogeneous PEs, the objective is to assign tensor elements across PEs to balance memory usage and minimize communication. In distributed memory systems, the non-uniform distribution of nonzeros across rows introduces storage imbalance, which must be carefully managed to ensure scalability and efficiency.

Let each row \( r_i \in R \) have an associated nonzero count \( \text{nnz}(r_i) \). The goal is to construct a partition \( \{R_1, R_2, \ldots, R_N\} \) of rows such that for every processing element \( \text{PE}_k \), the total number of nonzeros assigned satisfies:
\( \sum_{r_i \in R_k} \text{nnz}(r_i) \approx \frac{1}{N} \sum_{r_j \in R} \text{nnz}(r_j) \).
This ensures load balancing across PEs by equalizing the aggregate nonzero count rather than the number of rows but can lead to remote data accesses. \textit{Nexus Machine} exploits the resulting remote accesses using AMs: for remote access, idle PEs on the path may execute the associated AM payload, improving throughput and latency. 


\textit{Dissimilarity-Aware Mapping Strategy:} 
The primary tensor is partitioned using a dissimilarity-aware strategy~\cite{zed, gamma}, shown in Algorithm~\ref{algo:data_placement}, that spreads memory access and reduces contention, while secondary tensors are co-located or placed nearby to limit inter-PE traffic. Let \( L_i \subseteq \mathbb{M} \) denote memory banks accessed by row \( i \), and 
\(d(i,j) = |L_i \Delta L_j|.\)
The mapping: 
\(f_r: R \rightarrow \{\text{PE}_1, \ldots, \text{PE}_P\}\)
groups rows with similar \( L_i \) to the same PE and spreads dissimilar ones to avoid contention, enabling efficient AM execution.


\begin{algorithm}[t]
\footnotesize
\caption{Dissimilarity-Aware Data Partitioning}
\begin{algorithmic}[1]
\Require CSR tensor \( T \), number of PEs \( P \), memory banks \( \mathbb{M} \)
\Ensure Mapping \( f_r: R \rightarrow \{\text{PE}_1, \ldots, \text{PE}_P\} \)
\For{each row \( r_i \in R \)}
    \State \( L_i \gets \text{AccessedBanks}(r_i) \) \Comment{From column indices}
\EndFor
\For{each pair \( (r_i, r_j) \)}
    \State \( d(i,j) \gets |L_i \Delta L_j| \)   \Comment{Symmetric difference}
\EndFor
\State \( \{C_1, \ldots, C_P\} \gets \text{Cluster}(R, d(i,j)) \) \Comment{Group similar rows}
\For{each cluster \( C_k \)}
    \For{each \( r \in C_k \)}
        \State \( f_r(r) \gets \text{PE}_k \)  \Comment{Assign to PE}
    \EndFor
\EndFor
\State \Return \( f_r \)
\end{algorithmic}
\label{algo:data_placement}
\end{algorithm}
}

\textbf{Lightweight Runtime Manager.}
The \textit{Runtime Manager}, operating concurrently on the host processor, utilizes data placement information provided by the static compiler to generate a sequence of \textit{static AMs}.
These \textit{static AMs} are then loaded into the \textit{AM Queues} of individual PEs for execution.
Fig.~\ref{fig:message_format} illustrates the format of these compiler-generated \textit{static AMs}.
Each \textit{static AM} corresponds to a unique \textit{Op1} operation and is loaded onto the respective PE, aligned with its designated location.
Each \textit{static AM} contains the value of first operand, along with essential details such as the PE ID, and address for the dependent input operands and the result.
For every element in the first operand, the runtime manager generates a \textit{static AM} containing information about the operands and the result. 
The AM pairs this element, stored as \textit{Op1}, with the location of the second operand denoted by \textit{R1}, indicating the PE where it resides, and stores the local address as \textit{Op2}. 
Similarly, the result is stored at \textit{R2}, along with its local address as \textit{Result}.

\section{Evaluation Methodology}
\textit{Nexus Machine} is implemented in SystemVerilog using the parameters in Table~\ref{tab:arch_parameters}, and synthesized with Cadence Genus targeting a commercial 22nm FDSOI process. Memory blocks are instantiated via foundry-validated SRAM compilers for accurate area and timing characterization. The design is fully synthesizable and runs at up to 588 MHz. We develop a cycle-accurate Python simulator that models key behaviors including dynamic routing, congestion, and backpressure. This enables detailed analysis of dataflow and memory access patterns. Application workloads are compiled and mapped onto configuration memory and active message queues.

\begin{table}[h]
    \centering
    \resizebox{0.75\columnwidth}{!}{
    \begin{tabular}{ll}
        \toprule
        \textbf{Component} & \textbf{Configuration} \\ \midrule
        Array           & $4 \times 4$ INT16 array; \\
        SRAM            & 1KB per PE; 16KB Overall \\
        AM Queue        & 1KB FIFO with 70 bits per entry, 16KB Overall \\
        Main Memory     & 4.7GB/s, AXI4 interface \\
        \bottomrule
    \end{tabular}
    }
    \caption{\textit{Nexus Machine}'s architectural parameters.}
    \label{tab:arch_parameters}
\end{table}
\subsection{Baseline Architectures}
\begin{figure}[b!]
	\centering
	\includegraphics[width=\columnwidth]{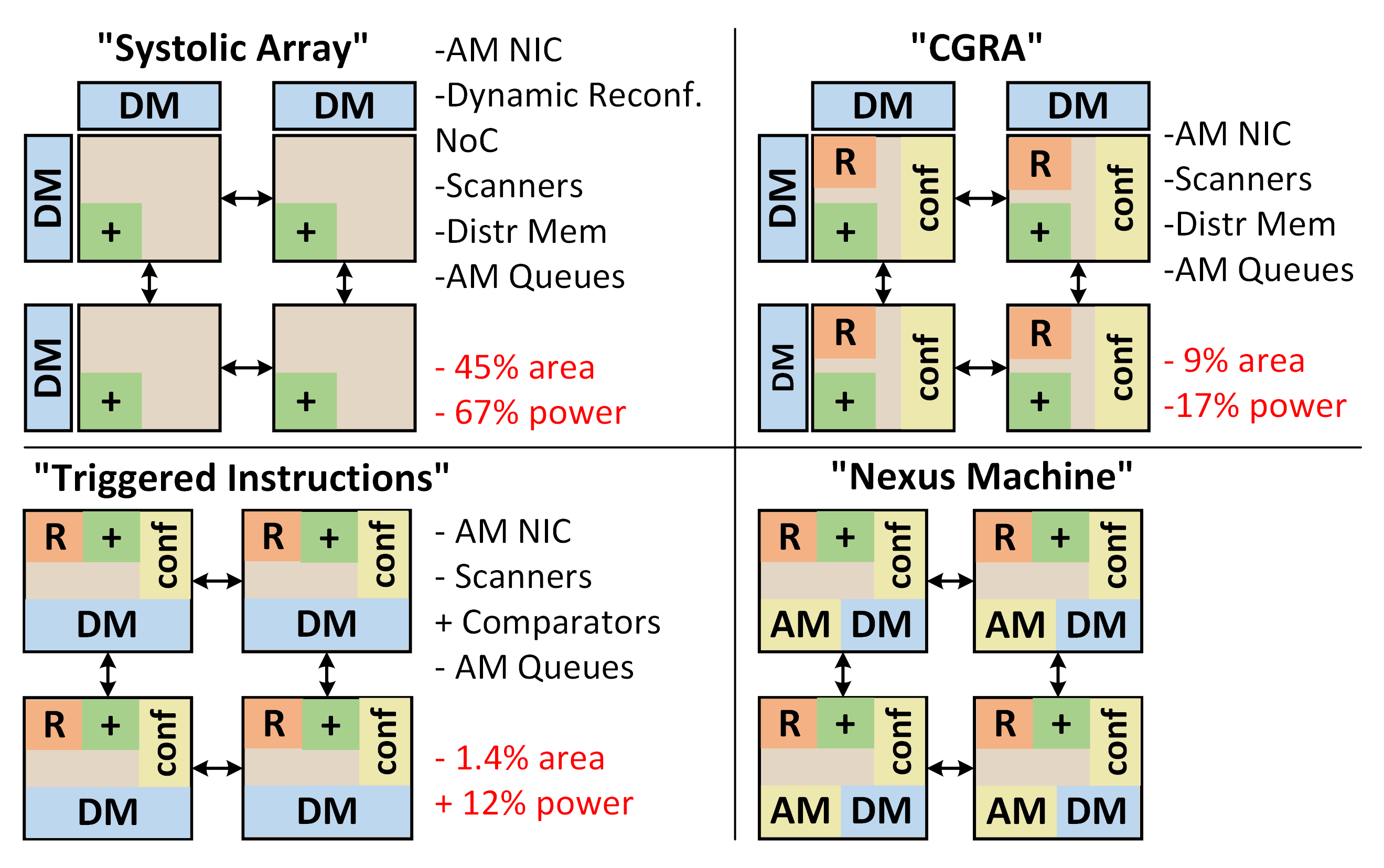}
	\caption{Ablation of \textit{Nexus Machine}'s features compared to its baselines} 
	\label{fig:ablation}
\end{figure}
We evaluate \textit{Nexus Machine} against four baseline architectures. For dense tensor acceleration, we use a \textbf{systolic array} modeled after the Google TPU~\cite{tpu_v4}. The \textbf{Generic CGRA} baseline is adapted from HyCube~\cite{hycube}, with eight memory banks along two edges to mitigate memory port limitations. We also evaluate the \textbf{TIA} baseline, based on the Triggered Instruction Architecture~\cite{tia}, and a \textbf{TIA-Valiant} variant that applies Valiant’s randomized minimal-path routing~\cite{valiant, valiant_minimal} as a traditional network load-balancing scheme.
All baselines are synthesized using the same technology node and provisioned with equal ALU counts for matched peak throughput. The \textit{TIA} baseline is evaluated using a cycle-accurate simulator, while CGRA designs use LLVM-based Morpher~\cite{morpher}, which models bank conflicts. To ensure fairness, each PE is allocated 2KB on-chip memory for all baselines, while \textit{Nexus Machine} uses 1KB SRAM and 1KB \textit{AM Queue} per PE. Memories are synthesized using the same SRAM compiler, with layout adapted per architecture as shown in Fig.~\ref{fig:ablation}: \textit{Nexus Machine} and \textit{TIA} use distributed memory, while \textit{systolic} and \textit{Generic CGRA} designs employ edge-based shared banks.

\subsection{Workloads}
\begin{figure*}[t!]
	\centering
	\includegraphics[width=0.95\textwidth]{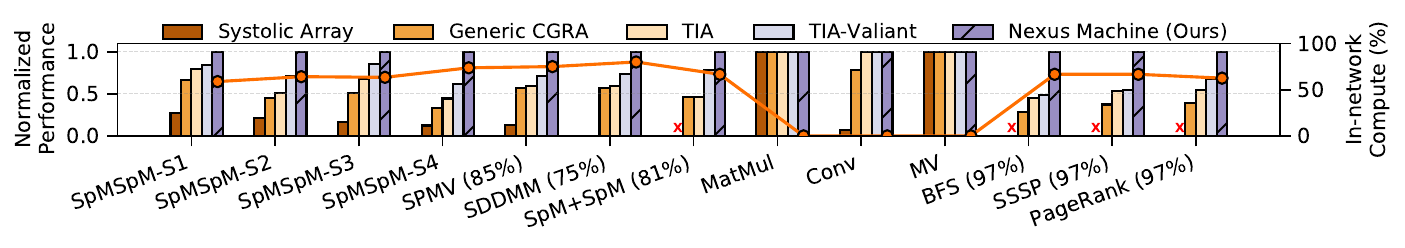}
	\caption{Normalized performance of \textit{Nexus Machine} relative to baselines. The right y-axis shows the percentage of computations performed in-network. Sparsity levels for each workload are noted in parentheses ().} 
	\label{fig:perf}
\end{figure*}
\begin{figure*}[t!]
	\centering
	\includegraphics[width=0.95\textwidth]{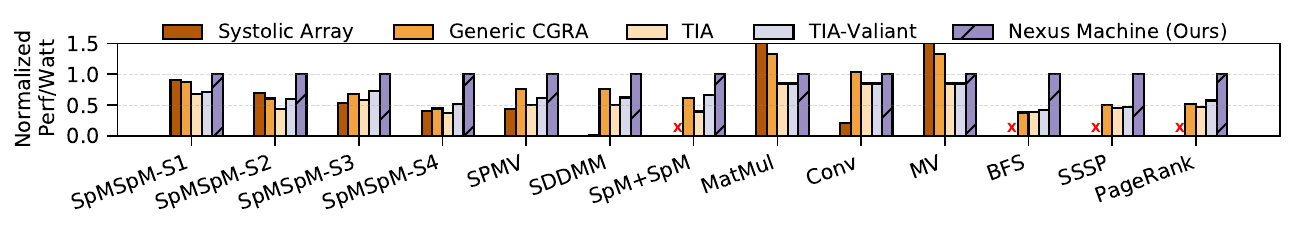}
	\caption{Normalized performance-per-watt of \textit{Nexus Machine} relative to baseline architectures.} 
	\label{fig:perf_watt}
\end{figure*}
\begin{figure}[t!]
	\centering
	\includegraphics[width=\columnwidth]{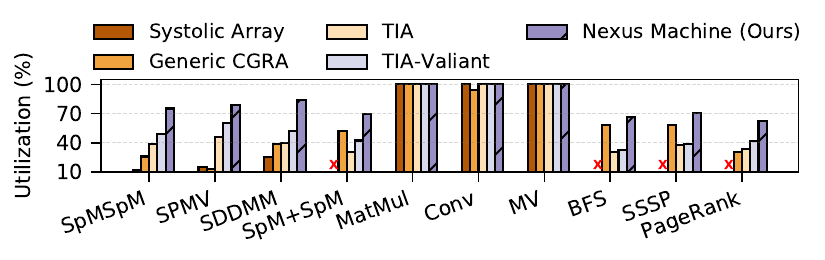}
	\caption{Fabric utilization (in \%) of \textit{Nexus Machine} relative to baseline architectures.} 
	\label{fig:utilization}
\end{figure}
\textit{Nexus Machine} is designed to accelerate workloads with irregular access patterns, including sparse tensor computations and graph processing. We evaluate \textit{Nexus Machine} across sparse, dense, and graph workloads to demonstrate architectural versatility.

\textbf{Sparse Matrix Multiplication (SpMSpM)} multiplies two sparse matrices in Compressed Sparse Row (CSR) format. We implement this using Gustavson's algorithm~\cite{gamma}. To study architectural behavior under varying sparsity, we apply sparsification methods~\cite{sanger, vitcod} to induce unstructured sparsity while controlling the sparsity-accuracy tradeoff. For stress testing, we evaluate sparsity levels up to 95\%, acknowledging that accuracy may degrade beyond 85\%. Workloads are categorized as: 
\textbf{S1}: both matrices moderately sparse (30–60\%), 
\textbf{S2}: A highly sparse (60–90\%), B moderately sparse, 
\textbf{S3}: A moderately sparse, B highly sparse, and 
\textbf{S4}: both highly sparse.


Beyond SpMSpM, we evaluate additional sparse workloads: 
\textbf{SpMV} multiplies a sparse matrix with a dense vector, 
\textbf{Sparse Matrix Addition (SpM+SpM)} performs element-wise addition of two CSR matrices, common in CNNs, 
\textbf{Sampled Dense-Dense Matrix Multiplication (SDDMM)} computes products only at sparse locations, useful in sparse attention and graph neural networks. 
For dense workloads, we execute 
\textbf{Matrix Multiplication (Matmul)}, 
\textbf{Matrix Vector Multiplication (MV)}, and 
\textbf{Convolution (Conv)}, all essential for CNNs and language models. For graph analytics, we evaluate 
\textbf{Breadth-First Search (BFS)}, 
\textbf{Single Source Shortest Path (SSSP)}, and 
\textbf{PageRank}, using adjacency list representations. 

Benchmarks are mapped to \textit{configuration memory} and \textit{AM queues} in \textit{Nexus Machine}. SpMv, SpMSpM, and SpM+SpM use a pruned and fine-tuned ResNet-50~\cite{resnet50}, with convolution layers converted into matrices~\cite{toeplitz}. SDDMM is evaluated using a sparse mask similar to ViTCoD~\cite{vitcod}, and dense workloads use unpruned ResNet-50. Graph workloads use infect-dublin~\cite{infect_dublin} dataset with graphs partitioned using Metis~\cite{metis} for balanced parallel execution.

\section{Evaluation Results}

Fig.~\ref{fig:perf}, Fig.~\ref{fig:perf_watt} and Fig.~\ref{fig:utilization} show that \textit{Nexus Machine} consistently delivers higher performance, power efficiency and fabric utilization over baseline architectures across workloads with varying sparsity. 

\subsection{Performance vs. Sparsity}
For Sparse workloads, \textit{Nexus Machine} consistently outperforms \textit{systolic}, and \textit{Generic CGRA} across varying sparsity regimes. For the SpMSpM workload, when the sparsity of the first tensor increases and the others remain fixed, performance decreases due to greater irregularity in AMs within the fabric. In contrast, increasing sparsity in other tensors while keeping the first fixed leads to improved performance, as AMs terminate early when they do not find corresponding elements in the other matrices, reducing computation time. The \textit{Generic CGRA} struggles to execute this kernel efficiently due to a higher number of bank conflicts because of irregular memory accesses and shared bank contention. Intuitively, the bank conflicts in \textit{Generic CGRA} roughly translate to network congestion for \textit{Nexus Machine}, alleviating the entire processing fabric from starving for compute. \textbf{The AM driven execution model dynamically adapts to data sparsity and minimizes redundant operations, ultimately enhancing performance under irregular computation patterns.}

For dense workloads, \textit{Nexus Machine}, \textit{TIA} and \textit{TIA-Valiant} baselines emulate systolic-style dataflow for MatMul and MV, achieving performance comparable to the \textit{systolic} and \textit{Generic CGRA} baselines. While the \textit{systolic} delivers the highest performance and best Perf/Watt for these workloads, it is inefficient for Conv due to im2col overhead~\cite{im2col} and cannot execute Conv natively. In contrast, \textit{Nexus Machine} efficiently handles Conv by replicating filters across PEs with minimal overhead. \textit{Generic CGRA} achieves near-optimal performance. \textbf{Despite being less efficient in Perf/Watt for dense workloads, \textit{Nexus Machine}’s reconfigurable architecture enables robust and versatile execution across diverse dense workloads.}

Finally, \textit{TIA} and \textit{TIA-Valiant} baselines, which lack any in-network computation capability, serve as ablation points to distinguish the benefits of en-route computation. The \textit{TIA-Valiant} variant routes to a randomized intermediate destination, distributing traffic evenly across the network, effectively reducing hotspot formations. As shown in Fig.~\ref{fig:perf}, \textit{Nexus Machine} performs a significantly greater portion of its computation on idle PEs en-route, leading to improved fabric utilization. \textbf{This showcases the architectural advantage of \textit{Nexus Machine} in exploiting in-network computation to enhance load balancing and overall system throughput.}

\begin{figure}[t!]
	\centering
	\includegraphics[width=\columnwidth]{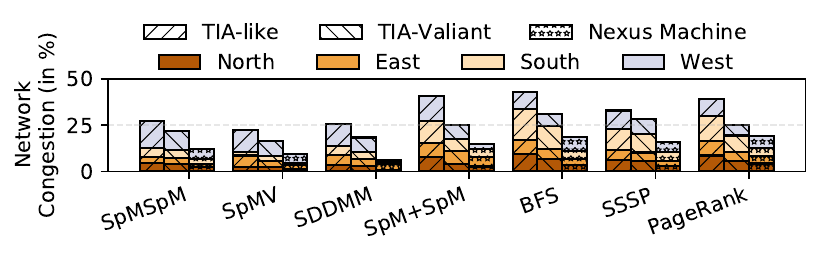}
	\caption{Network congestion at each input port of the \textit{Nexus Machine}, relative to \textit{TIA} baselines. Dense workloads are omitted due to minimal congestion from fixed dataflow patterns.} 
	\label{fig:congestion}
\end{figure}
Fig.~\ref{fig:congestion} shows network congestion across all input ports, where the \textit{Nexus Machine} reduces average congestion compared to the \textit{TIA} baseline due to its adaptive AM-based routing. Congestion data for \textit{Generic CGRA} is omitted due to its static routing determined at compile time, while \textit{systolic} uses fixed, deterministic routing. This distinction is reflected in compilation time: \textit{Generic CGRA} requires 7.22 seconds for static route resolution, while \textit{Nexus Machine} compiles in 0.55 seconds with runtime routing handled in hardware.

\vspace{-3mm}
\subsection{Overhead for AM inspired execution}
\begin{figure}[t!]
	\centering
    \includegraphics[width=1.05\columnwidth]{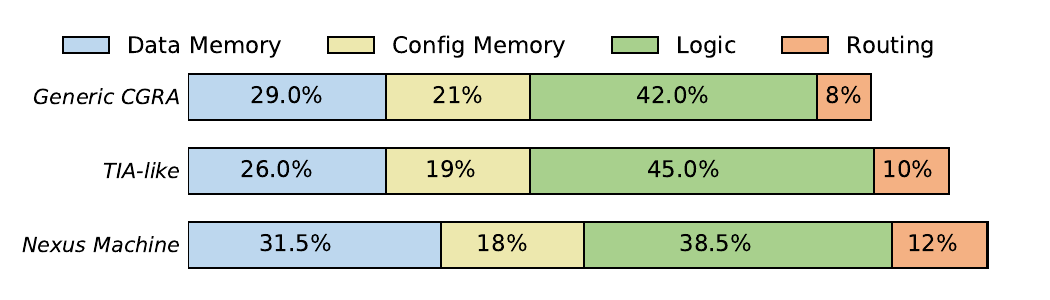}
	\caption{Area breakdown of the \textit{Nexus Machine} compared to \textit{Generic CGRA} and \textit{TIA} baselines.} 
	\label{fig:area}
\end{figure}

\textit{Generic CGRA} is used as the baseline to evaluate the impact of architectural enhancements in \textit{Nexus Machine}, specifically the inclusion of a dynamic NoC, \textit{AM NIC}, and distributed memories instead of the global data memories along the PE array edge for \textit{Generic CGRA}. The \textit{TIA} baseline isolates the contribution of the \textit{AM NIC}, its associated buffering structures and scanners.

\textbf{Area cost:} Fig.~\ref{fig:area} compares the area breakdown across \textit{Nexus Machine}, \textit{Generic CGRA}, and \textit{TIA}. \textit{Nexus Machine} exhibits 17.3\% and 5.2\% area overhead compared to \textit{Generic CGRA} and \textit{TIA}, respectively. Of this, 8\% is attributed to the \textit{AM queues} and related logic. An additional 3\% overhead over \textit{Generic CGRA} is due to the integration of scanner units. \textit{TIA} shows an 8\% increase relative to \textit{Generic CGRA} due to additional comparators. Both \textit{Nexus Machine} and \textit{TIA} incur a further 6\% overhead from dynamic routers implementing turn-model routing and congestion-aware control.

\textbf{Power Cost:} Fig.~\ref{fig:ablation} presents the power overhead of \textit{Nexus Machine} relative to the baselines. \textit{Nexus Machine} incurs a 17\% increase in total power compared to \textit{Generic CGRA}, with 8\% attributed to replicated configuration memories, 0.5\% to scanner units, 7\% to dynamic routers, and 6\% to control logic.
In comparison to the \textit{TIA} baseline, \textit{Nexus Machine} incurs 0.5\% additional power for scanners but benefits from a 12\% reduction in configuration memory power due to the absence of comparators, which contribute significantly to \textit{TIA}'s control logic overhead.

\textbf{These results indicate that despite \textit{Nexus Machine} introducing moderate area and power overheads, its design choices balance flexibility and efficiency by reducing reliance on high-cost control structures.}
\vspace{-3.5mm}
\subsection{Data Memory Size vs. Off-Chip Bandwidth}
\begin{figure}[t]
	\centering
	\includegraphics[width=\columnwidth]{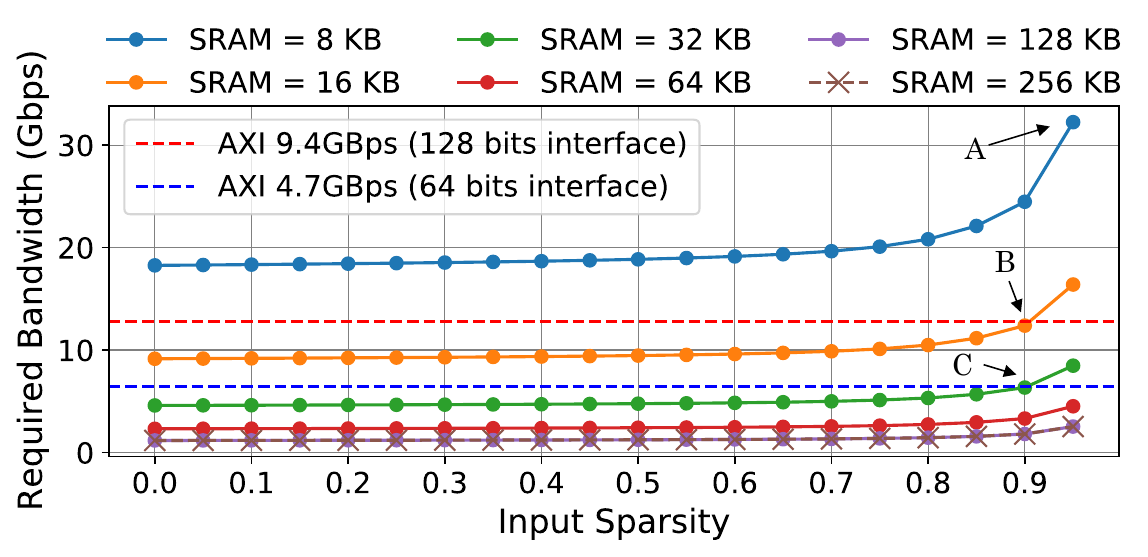}
	\caption{\textit{Nexus Machine} off-chip bandwidth required to achieve peak computational throughput for varying sparsity.} 
	\label{fig:dram_bandwidth}
\end{figure}
The efficiency of the architecture is affected by the tradeoffs between on-chip memory capacity and off-chip bandwidth, especially across workloads with varying sparsity. To study this relationship, we synthesize and characterize SpMSpM workloads at multiple sparsity levels, which directly affect computation per memory fetch. We apply the sparse tensor partitioning strategy, tiling tensors within on-chip SRAM to minimize off-chip transfers.

As shown in Fig.~\ref{fig:dram_bandwidth}, results reveal a key trade-off: as sparsity increases, computations per byte of off-chip traffic decreases, raising bandwidth requirements, even as \textit{Nexus Machine} sustains stable throughput. Beyond 256KB of on-chip SRAM, bandwidth usage stabilizes at its minimal achievable level. At extreme sparsity ($\approx$95\%), bandwidth usage rises up to 7× due to increased output movement, yet dense-equivalent throughput improves up to 16×, despite partial under-utilization.

\textit{Nexus Machine} was further evaluated using AXI4 interfaces configured at 16 beats with 64-bit and 128-bit widths. These experiments highlight design opportunities based on available AXI4 bandwidth. Design point A, with low on-chip memory requires a higher bandwidth. Conversely, our baseline Design Point B, with substantial on-chip memory, is well suited for workloads with low computation-to-transfer ratios. Lastly, Design Point C can reduce the budget of both on-chip and off-chip memory if the workloads have a higher compute intensity.

\subsection{\textit{Nexus Machine}'s Architecture Scalability}
\begin{figure}[t!]
	\centering
	\includegraphics[width=\columnwidth]{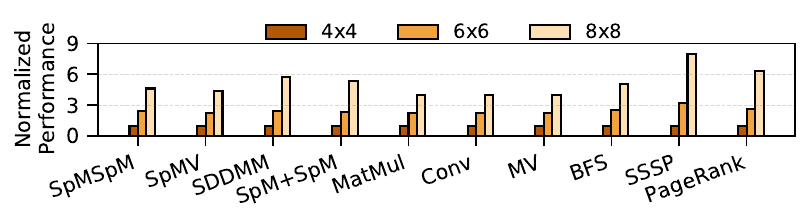}
	\caption{Normalized performance of the \textit{Nexus Machine} across increasing array sizes, illustrating its scalability with respect to architectural dimensions.} 
	\label{fig:scalability}
\end{figure}
Fig.~\ref{fig:scalability} evaluates the scalability of \textit{Nexus Machine} by comparing performance across different PE array sizes for various workloads. As the number of PEs increases, \textit{Nexus Machine} demonstrates close to linear scaling, showing its ability to scale efficiently. However, for irregular workloads, performance depends on the amount of computation per data element loaded. In such cases, more frequent data swapping, due to sparsity-related irregularity, introduces overhead and reduces overall fabric efficiency.

It shows fabric utilization across workloads, indicating that utilization is more sensitive to input sparsity than to problem size. \textbf{This confirms that \textit{Nexus Machine} maintains scalability across different problem scales, with performance primarily influenced by workload sparsity.}
\section{Related Works}
\subsection{CGRA support for irregularity} 
\label{section: related_works}

\textit{Generic CGRA} architectures are inefficient at handling irregular workloads. We classify prior studies proposing adaptations for irregularity and sparsity into server-scale solutions, addressing memory bandwidth and coarse task parallelism, and edge-device solutions that leverage finer-granularity parallelism with ultra-lightweight hardware mechanisms.

\textbf{Server-Scale:}
Fifer~\cite{fifer} enables irregular workloads on \textit{generic CGRAs} by decomposing them into regular stages, using queues to decouple and time-multiplex stages for better fabric utilization. However, it is inefficient for edge due to inter/intra-PE queues, frequent reconfiguration, and high data movement costs.
Capstan~\cite{capstan} features a checkerboard grid of compute and local memory units, with scanners and address generators for external memory. It uses sparse dataflow patterns and employs sparse-iteration techniques for efficiency and parallelism. Capstan requires additional hardware for data/metadata manipulation and expensive shuffle networks for PEs to access memory.
Extensor~\cite{extensor} employs hierarchical tensor intersections in an inner-product flow, exploiting data sparsity but sacrifices programmability, limiting its ability to handle diverse sparse tensor operations efficiently.
Dalorex~\cite{dalorex} uses a task-based model with uniform data distribution across tiles arranged in a 2D torus topology with a ruche network~\cite{ruche}. Each tile includes an in-order core, scratchpad, scheduler, and router. Parent tasks spawn new tasks into the NoC upon completion.
\textit{Nexus Machine} balances loads with in-network computation, using idle PEs en-route. While Dalorex works at task-level granularity, ideal for servers, \textit{Nexus Machine} offers instruction-level control for edge devices.

\textbf{Edge:}
Ultra-Elastic CGRA~\cite{uecgra} accelerates inter-iteration loop dependencies in irregular loops by exploiting dynamic voltage and frequency scaling (DVFS) at the granularity of PEs. However, its scope remains constrained by the range of supported workloads.
Stardust~\cite{stardust} and TACO~\cite{taco} provide compiler frameworks for Capstan-like architectures, managing data partitioning and transfers. Sparse Abstract Machine (SAM)\cite{sam} offers stream-dataflow abstractions with generic, composable sparse primitives, while SparseTIR\cite{sparsetir} explores composable abstractions for sparse deep learning. SAM's implementation Onyx~\cite{onyx} decouples tensor traversal from dataflows, enabling specialized optimizations as an intermediate representation by combining its primitives.
SPU~\cite{dgra} uses a stream dataflow model on a reconfigurable fabric composed of decomposable memories, switches, and compute units that split coarse-grained resources into finer-grained resources on the fabric. This enables flexibility in exploiting common stream-join and indirection data-dependence forms in irregular workloads.
Pipestitch~\cite{pipestitch} is a CGRA tailored for energy-efficient computation in sensor systems handling sparse ML and DSP tasks. 
It optimizes performance by converting annotated, parallelizable loops into ordered, pipelined dataflow threads.
The design incorporates a dispatch operator that utilizes specialized control flow PEs, forming a synchronization network for ordered-dataflow execution. 
The architecture demands more PEs compared to \textit{generic CGRAs}, while incurring higher data movement. Table~\ref{tab:sota_comparison} presents a comparison between \textit{Nexus Machine} and state-of-the-art edge CGRAs in terms of power, throughput, and power efficiency.
To ensure fairness, we put our best effort into comparing them, considering varied details among different sources.
\textit{Nexus Machine} showcases a better power efficiency of 194MOPS/mW compared to the SOTA CGRAs.

\begin{table}[t!]
    \centering
    \resizebox{\columnwidth}{!} {
    \begin{tabular}{|c|c|c|c|c|} \hline  
         &  \textbf{UE-CGRA}*&  \textbf{Pipestitch}*&\textbf{Triggered} & {\cellcolor[HTML]{C9C0BB}}{\textbf{Nexus}}\\ 
 & ~\cite{uecgra}& ~\cite{pipestitch}& \textbf{Instructions}**~\cite{tia}&{\cellcolor[HTML]{C9C0BB}}{\textbf{Machine}}\\ \hline  
         \textbf{Tech node}&  TSMC 28&  Sub-28&FDSOI 22& {\cellcolor[HTML]{C9C0BB}}{FDSOI 22}\\ \hline  
         \textbf{Setup}&  Post P\&R&  Post Synthesis&Post Synthesis& {\cellcolor[HTML]{C9C0BB}}{Post Synthesis}\\ \hline  
         \textbf{Peak Frequency Tested (MHz)}&  750&  50&588&{\cellcolor[HTML]{C9C0BB}}{588}\\ \hline
         \textbf{Power (mW)}&  14&  3.33&4.626& {\cellcolor[HTML]{C9C0BB}}{3.865}\\ \hline  
         \textbf{Peak Throughput}&  625&  558&490& {\cellcolor[HTML]{C9C0BB}}{748}\\ 
         \textbf{(MOPS)}& & &&{\cellcolor[HTML]{C9C0BB}}{}\\\hline  
         \textbf{Power Effici.}&  45&  167&106& {\cellcolor[HTML]{C9C0BB}}{194}\\  
         \textbf{(MOPS/mW)}& & &&{\cellcolor[HTML]{C9C0BB}}{}\\ \hline
    \end{tabular}
    }
    \begin{flushleft}
    \tiny{* Based on the latest values mentioned in Table III of RipTide~\cite{riptide} and approximations from Pipestitch~\cite{pipestitch} paper.\\
    ** Based on our implementation of the work.}
    \end{flushleft}
    \caption{\textit{Nexus Machine} compared to state-of-the-art CGRAs}
    \label{tab:sota_comparison}
\end{table}
\vspace{-0.25cm}
\subsection{Sparse Accelerators}

Recent literature introduces Sparse Matrix accelerators tailored for specific workloads.
SparTen~\cite{sparten} optimizes bitmap-based inner join operations for SpMM but requires offline load balancing.
EIE~\cite{eie} focuses on CNN fully connected layers but is constrained by on-chip storage. SpAtten~\cite{spatten}, for NLP, applies real-time quantization, ranking, and pruning in transformer attention layers.
SpArch~\cite{sparch} and OuterSPACE~\cite{outerspace} use outer-product dataflows for SpMSpM but face high partial-sum merging costs. MatRaptor~\cite{matraptor} and GAMMA~\cite{gamma} reduce these costs with row-wise product dataflows, and GAMMA adds specialized caching to reduce off-chip transfers.
Flexagon~\cite{flexagon} dynamically reconfigures dataflows for neural networks based on workload.
While these architectures are workload-specific, \textit{Nexus Machine} provides a generic CGRA for efficient execution of both irregular and regular workloads.
\vspace{-.1cm}
\section{Conclusion}
\textit{Nexus Machine} introduces a novel solution for enhancing the acceleration of irregular workloads. 
By enabling Active Messages within a reconfigurable architecture and computing on idle PEs in the network, it minimizes data movement and effectively mitigates load-balancing issues. 
This data-driven execution results in an average of 1.35x performance gain over prior state-of-the-art.
Overall, \textit{Nexus Machine} offers a promising architecture for resource-constrained edge devices, striking a balance between high performance, energy efficiency, and adaptability for irregular computational patterns.

\bibliographystyle{ACM-Reference-Format}
\bibliography{refs}

\end{document}